\documentclass[12pt]{article}
\pdfoutput=1
\usepackage{graphicx, fancyhdr, amssymb, amsmath, amsthm,
	epsfig, mathrsfs, verbatim,epsf}

\numberwithin{equation}{section}
\usepackage[section]{placeins}
\usepackage{natbib,enumerate,multicol}
\usepackage{graphicx, graphics, fancyhdr, fancyvrb, siunitx, epsf, verbatim, mathrsfs, amsmath, amssymb, amsthm, algorithmicx,algorithm, bm, algpseudocode, subcaption, mathrsfs, verbatim, epsf, siunitx, placeins, multicol, rotating, lscape, wrapfig, blkarray, setspace, enumitem, subcaption}
\usepackage[titletoc, title]{appendix}
\let\oldbibliography\thebibliography
\renewcommand{\thebibliography}[1]{\oldbibliography{#1}
	\setlength{\itemsep}{0pt}}

\newcommand{\E}[1]{\operatorname{E}\!\left[#1\right]}

\newcommand{\argmin}{\operatornamewithlimits{argmin}}
\newcommand{\utwi}[1]{\mbox{\boldmath $ #1$}}
\newcommand{\mLabel}[1]{\mbox{$\scriptstyle{#1}$}}
\newcommand{\Y}{{\utwi{Y}}}

\newcommand{\Z}{{\utwi{Z}}}
\newcommand{\PhiB}{{\utwi{\Phi}}}

\def\LL{\mathbf{L}}
\def\xx{{\bf x}} 
\def\r{{\bf r}}
\def\msfein{\mathrm{MSFE}_{\mathrm{in}}}
\def\tr{\mathrm{tr}}
\def\E{\mathbb E}
\def\P{\mathbb P}

\newtheorem{thm}{Theorem}
\newtheorem{lem}{Lemma}
\newtheorem{assumption}{Assumption}

\usepackage[explicit]{titlesec}

\titlespacing{\section}{0pt}{0pt}{0pt}

\title{High Dimensional Forecasting via Interpretable Vector Autoregression\footnote{The authors thank Gary Koop for providing his data
		transformation script.  This research was supported by an Amazon Web Services in
		Education Research Grant. IW was supported by  the European Union's Horizon 2020 research and innovation programme under the Marie Sk\l{}odowska-Curie grant agreement  No 832671., JB was supported by NSF DMS-1405746 and NSF DMS-1748166
		and DSM was supported by NSF DMS-1455172 and a Xerox Corporation Ltd. faculty research award.}}

\author{William B. Nicholson\footnote{Point72 Asset Management, L.P. Mr.\ Nicholson contributed to this article in his personal capacity.  The information, views, and opinions expressed herein are solely his own and do not necessarily represent the views of Point72.  Point72 is not responsible for, and did not verify for accuracy, any of the information contained herein.; email: wbn8@cornell.edu; url: http://www.wbnicholson.com},
	Ines Wilms\footnote{Assistant Professor, Department of Quantitative Economics, Maastricht Univeristy; email: i.wilms@maastrichtuniversity.nl; url:
		https://feb.kuleuven.be/ines.wilms}, 
	Jacob Bien\footnote{Assistant Professor, Department of Data Sciences and Operations, Marshall School of Business, University of Southern California; email: jbien@usc.edu; url:
		http://www-bcf.usc.edu/\~jbien/}, and David
	S. Matteson\footnote{Associate Professor, Department of Statistical
		Science and ILR School Department of Social Statistics, Cornell
		University, 1196 Comstock Hall, Ithaca, NY 14853; email:
		matteson@cornell.edu; url:
		http://stat.cornell.edu/\~matteson/}\\
}
\begin{document}
	
	\def\spacingset#1{\renewcommand{\baselinestretch}%
		{#1}\small\normalsize} \spacingset{1}
	
	\maketitle
	
	\begin{abstract}
Vector autoregression (VAR) is a fundamental tool for modeling multivariate time series.  However, as the number of component series is increased, the VAR model becomes overparameterized.  Several authors have addressed this issue by incorporating regularized approaches, such as the lasso in VAR estimation. Traditional approaches address overparameterization by selecting a low lag order, based on the assumption of short range dependence, assuming that a universal lag order applies to all components.  Such an approach constrains the relationship between the components and impedes forecast performance.  The lasso-based approaches perform much better in high-dimensional situations but do not incorporate the notion of lag order selection. 
We propose a new class of hierarchical lag structures (HLag) that embed the notion of lag selection into a convex regularizer. The key modeling tool is a group lasso with nested groups which guarantees that the sparsity pattern of lag coefficients honors the VAR's ordered structure.  The proposed HLag framework offers three basic structures, which allow for varying levels of flexibility, with many possible generalizations.  A simulation study demonstrates improved performance in forecasting and lag order selection over previous approaches, and macroeconomic, financial, and energy applications further highlight forecasting improvements as well as HLag's convenient, interpretable output. 
	\end{abstract}
	{\it Keywords:}   forecasting, group lasso, multivariate time series, variable selection, vector autoregression
	\newpage
	\spacingset{1.5} % Double spacing

%% INTRO %%
\section{Introduction}
\label{sec1}
Vector autoregression (VAR) has emerged as the standard-bearer for macroeconomic forecasting since the seminal work of \cite{sims1980}. VAR is also widely applied in numerous fields, including finance (e.g., \citealp{han15}), neuroscience (e.g, \citealp{hyvarinen}), and signal processing (e.g., \citealp{basu19}). 
The number of VAR parameters grows quadratically with the the number of component series, and, in the words of Sims, this ``profligate parameterization'' becomes intractable for large systems.
Without further assumptions, VAR modeling is infeasible except in limited situations with small number of components and  lag order.

Many approaches have been proposed for reducing the dimensionality of vector time series models, including 
canonical correlation analysis \citep{box1977canonical}, 
factor models (e.g., \citealp{forni2000generalized}, \citealp{stock2002forecasting}, \citealp{Bernanke05}), 
Bayesian models (e.g., \citealp{BGR}; \citealp{koop}), 
scalar component models  \citep{tiao1989model},
independent component analysis \citep{hyvarinen}, 
and dynamic orthogonal component models \citep{MattesonTsay2011}.
Recent approaches have  focused on imposing sparsity in the estimated coefficient matrices through the use of convex regularizers such as the lasso \citep{tibs}.  Most of these methods are, however, adapted from the standard regression setting and do not specifically leverage the ordered structure inherent to the lag coefficients in a VAR. 

This paper contributes to the lasso-based regularization literature on VAR estimation by proposing   a new class of regularized hierarchical lag structures (HLag), 
that embed lag order selection into a convex regularizer to simultaneously address the dimensionality and lag selection issues. 
HLag thus shifts the focus from obtaining estimates that are generally sparse (as measured by the number of nonzero autoregressive coefficients) to attaining estimates with {\em low maximal lag order}. 
As such, it combines several important advantages:
It produces interpretable models,
provides a flexible, computationally efficient method for lag order selection, and
offers practitioners the ability to fit VARs in situations where various components may have highly varying maximal lag orders.

Like other lasso-based methods, HLag methods have an interpretability 
advantage over factor and Bayesian models.
They provide direct insight into the series contributing to the forecasting of each  individual component. 
HLag has further exploratory uses relevant for the study of different economic applications, as we 
find our estimated models on the considered macroeconomic data sets to have an underlying economic interpretation. 
Comparable Bayesian methods, in contrast, primarily perform shrinkage making the estimated models more difficult to interpret, although they can be extended to include variable selection (e.g., stochastic search). 
Furthermore, factor models that are combinations of all the component series can greatly reduce dimensionality but forecast contributions from the original series are only implicit. By contrast, the sparse structure imposed by the HLag penalty explicitly identifies which components are contributing to model forecasts.

While our motivating goal is to produce interpretable models with improved point forecast performance, a convenient byproduct of the HLag framework is a flexible and computationally efficient method for \textit{lag order selection}.
Depending on the proposed HLag structure choice, each equation row in the VAR will either entirely truncate at a given lag (``componentwise HLag"), or allow the series's own lags to truncate at a different order than those of other series (``own/other HLag"), or allow every (cross) component series  to have its own lag order (``elementwise HLag").  Such lag structures are conveniently depicted in a ``Maxlag matrix''  which we introduce and use throughout the paper.

Furthermore,  HLag penalties are unique in providing a computationally tractable way to fit high order VARs, i.e., those with a \textit{large}  maximal lag order  $(pmax)$. They allow the possibility of certain components requiring large max-lag orders without having to enumerate over all combinations of choices. Practitioners, however,  typically choose a relatively small $pmax$. We believe that this practice is in part due to the limitations of current methods:
information criteria make it impossible to estimate VARs with large $pmax$ by least squares as the number of candidate lag orders scales exponentially with the number of components $k$.  Not only is it computationally demanding to estimate so many models, overfitting  also becomes a concern. 
Likewise, traditional lasso VAR forecasting performance degrades when  $pmax$ is too large,
and many Bayesian approaches, while statistically viable, are computationally infeasible or prohibitive, 
as we  will illustrate through simulations and applications.

In Section \ref{Sec2new} we review the literature on dimension reduction methods to address the VAR's overparametrization problem.
In Section \ref{Sec2} we introduce the HLag framework.  
The three aforementioned hierarchical lag structures are proposed in Section \ref{sec:lag-struct-hier}.  As detailed above, these structures vary in the degree to which lag order selection is common across different components.  For each lag structure, a corresponding HLag model is  detailed in Section \ref{sec:lag-hier-group} for attaining that sparsity structure.
Theoretical properties of high-dimensional VARs estimated by HLag are analyzed in Section \ref{subsec:theory}. 
The proposed methodology allows for flexible estimation in high dimensional settings with a single tuning parameter.
We develop algorithms in Section \ref{Sec3} that are computationally efficient and parallelizable across components.
Simulations in Section \ref{Sec4} and  
applications in Section \ref{Sec5} 
highlight HLag's advantages in forecasting and lag order selection.  

%% LITERATURE %%
\section{Review of Mitigating VAR Overparametrization \label{Sec2new}}
We summarize the most popular approaches to address the VAR's overparametrization problem and  discuss their link to the HLag framework. 

\subsection{Information Criteria}
Traditional approaches address overparametrization by selecting a low lag order. Early attempts utilize least squares estimation with an information criterion or hypothesis testing \citep{lutk1}. The asymptotic theory of these approaches is well developed in the fixed-dimensional setting, in which the time series length $T$ grows while the number of components $k$ and maximal lag order $pmax$ are held fixed \citep{white2001asymptotic}. However, for small $T$, it has been observed that no criterion works well \citep{nick}.
\cite{Gonz} find that for fixed $k$ and $pmax$, when
$T$ is relatively small, Akaike's Information Criterion (AIC) tends to overfit whereas Schwarz's Information Criterion (BIC)  tends to severely underfit. 
Despite their shortcomings, AIC, BIC, and corrected AIC (\citealt{Hurvich89}) are still the preferred lag order selection tools by most practitioners 
\citep{Lutk, tsay2013multivariate}.

A drawback with such approaches is, however, that they typically require the strong assumption of a single, universal lag order that applies across all components. 
While this reduces the computational complexity of model selection, it has little statistical or economic justification, unnecessarily
constrains the dynamic relationship between the components, and impedes forecast performance.   An important motivating goal of the HLag framework is to relax this strong assumption.
\cite{karlsson} show that violation of the universal lag order assumption can lead to overparameterized models or the imposition of false zero restrictions.  They instead suggest considering \emph{componentwise} specifications that allow each marginal regression to have a different lag order (sometimes referred to as an \emph{asymmetric VAR}).    
One such procedure \citep{hsiao} starts from univariate autoregressions and sequentially adds lagged components according to Akaike's ``Final Prediction Error'' \citep{akaike}.
However, this requires an \emph{a priori} ranking of components based on their perceived predictive power, which is inherently subjective.
\cite{Keating} offers a more general method which estimates all potential $pmax^{k}$ componentwise VARs and utilizes AIC/BIC for lag order selection.  Such an approach is computationally intractable and standard asymptotic justifications are inapplicable if the number of components $k$ is large.  
\cite{Ding} present several specifications which allow for varying lag order within a Bayesian framework.
Markov chain Monte Carlo estimation methods with spike and slab priors are proposed, but these are computationally intensive, and estimation becomes intractable in high dimensions though  recent advances have been made by \cite{Primiceri17}. 

Given the difficulties with lag order selection in VARs, many authors have turned instead to shrinkage-based approaches, which impose sparsity, or other economically-motivated restrictions, on the parameter space to make reliable estimation tractable, and are discussed below.

\subsection{Bayesian Shrinkage} 
Early shrinkage methods, such as \cite{Litterman1979}, take a pragmatic Bayesian perspective. Many  of them (e.g., \citealp{BGR}; \citealp{koop}) apply the \emph{Minnesota prior,} which uses natural conjugate priors to shrink the VAR toward either an intercept-only model or  a vector random walk, depending on the context.
The prior covariance is specified so as to incorporate the belief that a series' \emph{own} lags are more informative than \emph{other} lags and that lower lags are more informative than higher lags.  
With this prior structure, coefficients at high lags will have a prior mean of zero and a prior variance that decays with the lag.  Hence, coefficients with higher lags are shrunk more toward zero. However, unlike the HLag methods but similar to ridge regression, coefficients will not be estimated as exactly zero.

The own/other HLag penalty proposed below is inspired by this Minnesota prior.  It also has the propensity to prioritize own lags over other lags and to assign a greater penalty to distant lags, but it formalizes these relationships by embedding two layers of hierarchy into a convex regularization framework. One layer (within each lag vector) prioritizes own lags before other lags. Another layer (across lag vectors) penalizes distant lags more than recent lags since the former can only be included in the model if the latter are selected.

The Bayesian literature on dealing with overparametrization of VARs is rapidly growing, with many recent advances on, amongst others, improved prior choices (e.g., \citealp{carriero2012}, \citealp{GLP}), 
stochastic volatility (e.g., \citealp{Carriero17}), time-varying parameter estimation (e.g., \citealp{koop2013}), and dimension reduction via compressing (\citealp{koop2017}). 

\subsection{Factor Models} 
Factor models form another widely used class to overcome the VAR's overparameterization and have been used extensively for macroeconomic forecasting (e.g., \citealp{stock2002forecasting}). Here,
the factors serve the purpose of dimension reduction since the  information contained in the original high dimensional data set is summarized---often using principal component analysis---in a small number of factors. 
While Factor Augmented VARs (FAVAR) (e.g., \citealp{Bernanke05}) include one or more factors in addition to the observables, all observables are expressed as a weighted average of factors in Dynamic Factor Models (e.g., \citealp{forni2000generalized}). 

\subsection{ Lasso-based Regularization} 
Other shrinkage approaches have incorporated the lasso \citep{tibs}.  \cite{Hsu} consider the lasso with common information criterion methods for model selection.  The use of the lasso mitigates the need to conduct an exhaustive search over the space of all $2^{k^2pmax}$ possible models but does not explicitly encourage lags to be small. 
HLag, in contrast, forces low lag coefficients to be selected before corresponding high lag coefficients, thereby specifically shrinking toward low lag order solutions. As will be illustrated through simulations and empirical applications, this   often improves forecast performance. 

To account for the VAR's inherent ordered structure, \cite{lozano2009grouped} use a group lasso \citep{yuan} penalty to group together coefficients within a common component. 
\cite{BickelSong} 
treat each variable's own
lags differently from other variables' lags (similar to the  own/other Hlag penalty we propose), 
consider a group lasso structure and additionally down-weight higher lags via scaling the penalty parameter by an increasing function of the coefficients' lag.
The authors note that the functional form of these weights is arbitrary, but the estimates are sensitive to the choice of weights.  A similar \emph{truncating lasso} penalty is proposed by \cite{shojaie2010discovering} and refined by \cite{shojaie2012adaptive} in the context of graphical Granger causality. However, unlike HLag, this framework requires a functional form assumption on the decay of the weights as well as a
two-dimensional penalty parameter search which generally squares the computational burden.

%% METHODOLOGY %%
\section{Methodology\label{Sec2} }
Let $\{\mathbf{y}_t\in\mathbb R^k\}_{t=1}^T$ denote a $k$-dimensional vector time series of length $T$.
A $p$th order vector autoregression $\text{VAR}_k(p)$ may be expressed as a multivariate regression
\begin{align}
\label{VAR1a}
\mathbf{y}_t=\utwi{\nu}+\PhiB^{(1)}\mathbf{y}_{t-1}+\dots+\PhiB^{(p)}\mathbf{y}_{t-p}+\mathbf{u}_t, \; \text{ for } \;t=1,\ldots,T,
\end{align}
conditional on initial values $\{\mathbf{y}_{-(p-1)},\ldots, \mathbf{y}_{0}\}$, 
where $\utwi{\nu}\in\mathbb R^k$ denotes an intercept vector,  
$\{ \PhiB^{(\ell)} \in \mathbb R^{k \times k} \}_{\ell=1}^p$
are lag-$\ell$ coefficient matrices,
and 
$\{\mathbf{u}_t\in\mathbb R^k\}_{t=1}^T$ is a mean zero white
noise  vector time series with unspecified
$k \times k$ nonsingular contemporaneous covariance matrix $\mathbf{\Sigma}_u$.

In the classical low-dimensional setting in which $T > kp$, one may
perform least squares to fit the $\text{VAR}_k(p)$ model, minimizing
\begin{align}
\sum_{t=1}^T\|\mathbf{y}_t-\utwi{\nu}-\sum_{\ell=1}^p\PhiB^{(\ell)}\mathbf{y}_{t-\ell}\|_2^2\label{eq:ls}
\end{align}
over $\boldsymbol\nu$ and $\{\PhiB^{(\ell)}\}$, where $\|{\bf a}\|_2=(\sum_i {\bf a}_i^2)^{1/2}$ denotes the Euclidean
norm of a vector $\bf a$.
We will find it convenient to express the VAR using compact matrix notation:

\begin{align}
&   \begin{array}[r]{lclc}
\mathbf{Y}=[\mathbf{y}_1~\cdots~\mathbf{y}_T]& (k\times T); & \qquad   \PhiB=[\PhiB^{(1)}~\cdots~\PhiB^{(p)}] & (k\times kp);\\
\mathbf{z}_t=[\mathbf{y}_{t-1}^\top~\cdots~\mathbf{y}_{t-p}^\top]^{\top} & (kp\times 1); & \qquad\mathbf{Z}=[\mathbf{z}_1~\cdots~\mathbf{z}_{T}] & (kp\times T);\\
\mathbf{U}=[\mathbf{u}_1~\cdots~\mathbf{u}_T] & (k\times T); & \qquad \utwi{1}=[1~\cdots~1]^{\top} & (T \times 1).
\end{array} \label{notation}
\end{align}
Equation \eqref{VAR1a} is then simply   
\begin{align*}
\mathbf{Y}=\utwi{\nu}\utwi{1}^{\top}+\PhiB\mathbf{Z}+\mathbf{U},
\end{align*}
and the least squares procedure \eqref{eq:ls} can be expressed as
minimizing
$$
\|\mathbf{Y}-\utwi{\nu}\utwi{1}^{\top}-\PhiB\mathbf{Z}\|_2^2
$$
over $\boldsymbol\nu$ and $\PhiB$, where $\|\mathbf A\|_2$ denotes the Frobenius norm of the matrix
$\mathbf A$, that is the Euclidean norm of $\text{vec}(\mathbf A)$
(not to be mistaken for the operator norm, which does not
appear in this paper).

Estimating the parameters of this model is challenging unless $T$ is
sufficiently large. Indeed, when $T > kp$ but $kp/T\approx 1$, estimation by least squares becomes imprecise. We therefore seek to incorporate reasonable structural assumptions on the parameter space to make estimation tractable for moderate to small $T$.  Multiple authors have considered using the lasso penalty, building in the assumption that the lagged coefficient matrices $\PhiB^{(\ell)}$ are sparse (e.g., \citealp{BickelSong,Davis,Hsu}); theoretical work has elucidated how such structural assumptions can lead to better estimation performance even when the number of parameters is large (e.g., \citealp{basu2013estimation}, \citealp{Melnyk16}, \citealp{lin17}).  In what follows, we define a class of sparsity patterns, which we call hierarchical lag or HLag structures, that arises in the context of multivariate time series.

\subsection{HLag: Hierarchical Lag Structures}
\label{sec:lag-struct-hier}
In Equation \eqref{VAR1a}, the parameter $\PhiB_{ij}^{(\ell)}$ controls the dynamic dependence of the $i$th component of $\mathbf{y}_t$ on the $j$th component of $\mathbf{y}_{t-\ell}$.  In describing HLag structures, 
we will use the following notational convention:  for $1\le\ell\le p$, let
\begin{align*}
\PhiB^{(\ell:p)}&=[\PhiB^{(\ell)}~\cdots~\PhiB^{(p)}]\in\mathbb
R^{k\times k(p-\ell+1)}\\
\PhiB_{i}^{(\ell:p)}&=[\PhiB_i^{(\ell)}~\cdots~\PhiB_i^{(p)}]
\in\mathbb R^{1\times k(p-\ell+1)}\\
\PhiB_{ij}^{(\ell:p)}&=[\PhiB_{ij}^{(\ell)}~\cdots~\PhiB_{ij}^{(p)}]\in\mathbb
R^{1\times (p-\ell+1)}.
\end{align*}
Consider the $k\times k$ matrix of \emph{elementwise} coefficient lags $\mathbf L$ defined by
\begin{equation}
\LL_{ij}=\max\{\ell :\PhiB_{ij}^{(\ell)}\neq0\}, \nonumber
\end{equation}
in which we define $\LL_{ij}=0$ if $\PhiB_{ij}^{(\ell)}=0$ for all $\ell=1,\ldots,p$. 
Therefore, each $\LL_{ij}$ denotes the maximal coefficient lag (maxlag) for component $j$ in the regression model for component $i$.
In particular, $\LL_{ij}$ is the smallest $\ell$ such that
$\PhiB_{ij}^{([\ell+1]:p)}=\boldsymbol{0}$.  Note that the maxlag matrix $\LL$
is not symmetric, in general.  There are numerous HLag
structures that one can consider within the context of the
$\text{VAR}_k(p)$ model.  The simplest such structure is that
$\LL_{ij}=L$ for all $i$ and $j$, meaning that there is a
\emph{universal} (U) maxlag
that is shared by every pair of components.  Expressed in terms of Equation \eqref{VAR1a}, this would say that $\PhiB^{([L+1]:p)}=\boldsymbol{0}$ and that $\PhiB^{(L)}_{ij}\neq0$ for all $1\le i,j\le k$.  While the methodology we introduce
can be easily extended to this and many other potential HLag structures, in this paper we focus on the following three fundamental  structures.

\begin{enumerate}
	\item {\bf Componentwise (C).}  
	A componentwise HLag structure allows each of the $k$ marginal equations from \eqref{VAR1a} to have its own maxlag, but all components within each equation must share the same maximal lag: 
	$$\LL_{ij}=L_i \;\; \forall j, \;\; \mathrm{for} \;\; i = 1,\ldots k.$$ 
	Hence in Equation \eqref{VAR1a}, this implies $\PhiB^{([L_i+1]:p)}_{i}=\boldsymbol{0}$ and $\PhiB^{(L_i)}_{ij}\neq0$ for all $i$ and $j$.  This componentwise HLag active set structure (shaded) is illustrated in Figure \ref{fig:fig1a}.
	
	\begin{figure}
		\centering
		
		\begin{minipage}{0.6\linewidth}
			\includegraphics[width=1\linewidth]{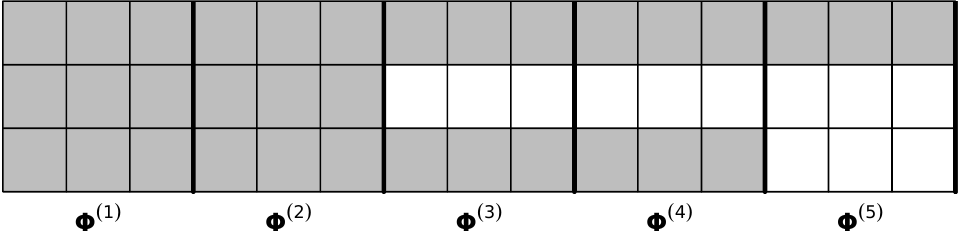} 
		\end{minipage}
		\begin{minipage}{0.3\linewidth}
			\vspace{-0.5cm}\large$$
			\LL^C =\begin{pmatrix}
			5&5&5\\
			2&2&2\\
			4&4&4
			\end{pmatrix}
			$$
		\end{minipage}
		\caption{A componentwise (C) HLag active set structure (shaded):
			$\text{HLag}^C_3(5)$.}\label{fig:fig1a}
		\vspace{0.25cm}
		
		\begin{minipage}{0.6\linewidth}
			\includegraphics[width=1.0\linewidth]{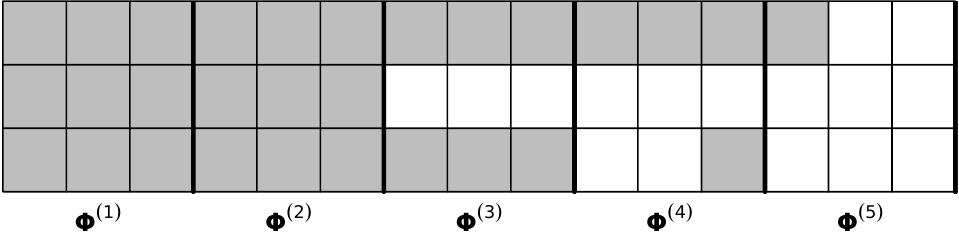} 
		\end{minipage}
		\begin{minipage}{0.3\linewidth}
			\vspace{-0.5cm}\large$$
			\LL^O =\begin{pmatrix}
			5&4&4\\
			2&2&2\\
			3&3&4
			\end{pmatrix}
			$$
		\end{minipage}
		\caption{An own-other (O) HLag active set structure (shaded): $\text{HLag}^O_3(5)$.}\label{fig:fig1b}
		\vspace{0.25cm}
		
		\begin{minipage}{0.6\linewidth}
			\includegraphics[width=1.0\linewidth]{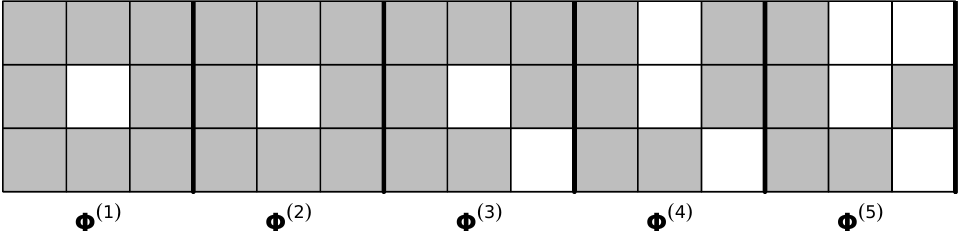} 
		\end{minipage}
		\begin{minipage}{0.3\linewidth}
			\vspace{-0.5cm}\large$$
			\LL ^E=\begin{pmatrix}
			5&3&4\\
			5&0&5\\
			5&5&2
			\end{pmatrix}
			$$
		\end{minipage}
		\caption{An elementwise (E) HLag active set structure (shaded): $\text{HLag}^E_3(5)$.}\label{fig:fig1c}
	\end{figure}
	
	\item {\bf Own-Other (O).} The own-other  HLag structure 
	is similar to the componentwise one, 
	but with an added within-lag hierarchy that imposes the mild assumption that a series' own lags $(i=j)$ are more informative than other lags $(i\neq j)$.  Thus, diagonal elements are prioritized before off-diagonal elements within each lag, componentwise (i.e., row-wise).  
	In particular, 
	$$\LL_{ij}=L_i^{other} \;\; \mathrm{for} \;\;  i\neq j \;\; \mathrm{and} \;\; \LL_{ii}\in\{L_i^{other},L_i^{other}+1\}, \;\; \mathrm{for} \;\; i = 1,\ldots k.$$ 
	This HLag structure allows each component of ${\bf y}_t$ to have longer range lagged 
	self-dependence than lagged cross-dependencies. 
	This own-other HLag structure is illustrated in Figure \ref{fig:fig1b}.
	
	\item {\bf Elementwise (E).} Finally, we consider a completely flexible structure in which the elements of $\LL$ have no stipulated relationships.  
	Figure \ref{fig:fig1c} illustrates this elementwise HLag structure.
\end{enumerate}

In the next section, we introduce the proposed class of HLag estimators aimed at estimating $\text{VAR}_k(p)$ models 
while shrinking the elements of $\LL$ towards zero by incorporating the three HLag structures described above.

\subsection{HLag: Hierarchical Group Lasso for Lag Structured VARs}
\label{sec:lag-hier-group}
In this section, we introduce convex penalties specifically tailored for attaining the three lag structures presented in the previous section.  Our primary modeling tool is the hierarchical group lasso \citep{Zhao09, Yan2017}, which is a group lasso \citep{yuan} with a nested group structure.  The group lasso is a sum of (unsquared) Euclidean norms and is used in statistical modeling as a penalty to encourage groups of parameters to be set to zero simultaneously.  Using nested groups leads to hierarchical sparsity constraints in which one set of parameters being zero implies that another set is also zero.  This penalty has been applied to multiple statistical problems including regression models with interactions \citep{Zhao09,Jenatton10,Radchenko10,Bach12,Bien13,lim2013learning,haris2014convex,She2014}, covariance estimation \citep{bien2016convex}, additive modeling \citep{lou2014sparse}, and time series \citep{suo2014ordered}.  This last work focuses on transfer function estimation, in this case scalar regression with multiple time-lagged covariates whose coefficients decay with lag.

For each hierarchical lag structure presented above, we propose an estimator based on a convex optimization problem:
\begin{align}
\label{optim}
\min_{\boldsymbol{\nu},\PhiB} \left\{\frac{1}{2T}\|\mathbf{Y}-\utwi{\nu}\mathbf{1}^{\top}-\PhiB\mathbf{Z}\|_2^2+\lambda {\cal P}^{}_{\text{HLag}}(\PhiB)\right\},
\end{align}
in which ${\cal P}^{}_{\text{HLag}}$ denotes a hierarchical lag group (HLag) penalty function. We propose three such penalty functions: componentwise; own-other; and elementwise; and discuss their relative merits. 

\begin{enumerate}
	\item {\bf HLag$^{C}$} aims for a {\em componentwise} hierarchical lag structure and is defined by
	\begin{align}
	\label{OV1}
	{\cal P}^C_{\text{HLag}}({\bf \PhiB})= \sum_{i=1}^k\sum_{\ell=1}^p\|\PhiB_i^{(\ell:p)}\|_2 ,
	\end{align}
	in which $\|\mathbf A\|_2$ denotes the Euclidean norm of $\text{vec}(\mathbf A)$, for a matrix $\mathbf A$.
	As the penalty parameter $\lambda \ge  0$ is increased, we have $\hat{\PhiB}_i^{(\ell:p)}=\boldsymbol{0}$ for more $i$, and for smaller $\ell$.  This componentwise HLag structure builds in the condition that if $\hat\PhiB_{i}^{(\ell)}=0$, then $\hat\PhiB_{i}^{(\ell')}=0$ for all $\ell'>\ell$, for each $i = 1,\ldots,k$. This structure favors lower maxlag models componentwise, rather than simply giving sparse $\PhiB$ estimates with no particular structure.

	\item {\bf HLag$^{O}$} aims for a {\em own-other} hierarchical lag structure and is defined by
	\begin{align}
	{\cal P}^O_{\text{HLag}}({\bf \PhiB})=\sum_{i=1}^k\sum_{\ell=1}^p\left[\|\PhiB_{i}^{(\ell:p)}\|_2+\|(\PhiB_{i,-i}^{(\ell)}, \PhiB_{i}^{([\ell+1]:p)})\|_2\right],
	\label{OV2}
	\end{align}
	in which $\PhiB_{i,-i}^{(\ell)}=\{\PhiB_{ij}^{(\ell)}:j\neq i\}$, and where we adopt the convention that $\PhiB_{i}^{([p+1]:p)}=\boldsymbol{0}$.
	The first term in this penalty is identical to that of \eqref{OV1}.  The difference is the addition of the second penalty term, which is just like the first except that it omits $\PhiB_{ii}^{(\ell)}$.  This penalty allows sparsity patterns in which the influence of component $i$ on itself may be nonzero at lag $\ell$ even though the influence of other components is thought to be zero at that lag. 
	This model ensures that, for all $\ell'>\ell$,
	$\hat\PhiB_{i}^{(\ell)}=\boldsymbol{0}$ implies $\hat\PhiB_{i}^{(\ell')}=\boldsymbol{0}$ and $\hat\PhiB_{ii}^{(\ell)}=\boldsymbol{0}$ implies $\hat\PhiB_{i,-i}^{(\ell'+1)}=\boldsymbol{0}$.
	This accomplishes the desired own-other HLag structure such that  $\LL_{i,-i}=L_i^{other}{\bf 1}_{k-1}$ and $\LL_{ii}\in\{L_i^{other},L_i^{other}+1\}$, componentwise.
	
	\item {\bf HLag$^{E}$} aims for an {\em elementwise} hierarchical lag structure and is defined by
	\begin{align}
	\label{OV3}
	{\cal P}^E_{\text{HLag}}({\bf \PhiB})=\sum_{i=1}^k\sum_{j=1}^k\sum_{\ell=1}^p\|\PhiB_{ij}^{(\ell:p)}\|_2.
	\end{align}
	Here, each of the $k^2$ pairs of components can have its own maxlag, such that $\PhiB_{ij}^{(\ell:p)}=\boldsymbol{0}$ may occur for different values of $\ell$ for each pair $i$ and $j$.  While this model is the most flexible of the three, it also borrows the least strength across the different components.  When $\LL_{ij}$ differ for all $i$ and $j$, we expect this method to do well, whereas when, for example $\LL_{ij}=L_i$, we expect it to be inefficient relative to \eqref{OV1}.
\end{enumerate}
Since all three penalty functions are based on hierarchical group lasso penalties, a unified computational approach to solve each is detailed in Section \ref{Sec3}.
First, we discuss theoretical properties of  HLag.

\subsection{Theoretical Properties} \label{subsec:theory}
We build on \cite{basu2013estimation} to analyze theoretical properties of
high-dimensional VARs estimated by HLag.
Consider a fixed realization of $\{\mathbf{y}_t\}_{t = -(p-1)}^{T}$  generated from the VAR model \eqref{VAR1a}
with fixed autoregressive order $p$ and $\mathbf{u}_t \overset{iid}{\sim} N({\bf 0}, \mathbf{\Sigma}_u)$. 
Denote the corresponding true maxlag matrix by ${\bf L}$. We make the following assumptions.

\begin{assumption}\label{assumptions}
	The VAR model is stable, such that $\text{det}\{\PhiB(z)\} \neq 0$ for all 
	$\{ z \in  \mathbb{C} : |z|\leq 1 \}$, where
	$${\PhiB}(z)  = {\bf I} -  \PhiB^{(1)}z -  \PhiB^{(2)}z^2 - \ldots -  \PhiB^{(p)}z^p;$$ and the error covariance matrix $\mathbf{\Sigma}_u$ is positive definite such that its minimum eigenvalue $\Lambda_{\text{min}}(\mathbf{\Sigma}_u)>0$ and its maximum eigenvalue  $\Lambda_{\text{max}}(\mathbf{\Sigma}_u)<\infty$. 
\end{assumption}

These assumptions are standard in the time series literature. 
Define the following two measures of stability of the VAR process, which will be useful for our theoretical analysis (see \citealp{basu2013estimation} for more detail)
$$
\mu_{\text{min}}(\PhiB) = \min_{|z|=1}\Lambda_{\text{min}}(\PhiB^*(z)\PhiB(z)), \ \text{and} \
\mu_{\text{max}}(\PhiB) = \max_{|z|=1}\Lambda_{\text{max}}(\PhiB^*(z)\PhiB(z)),
$$
where $\PhiB^*(\cdot)$ denotes the conjugate transpose of a complex matrix. 

We derive a bound on the in-sample prediction error. Define the in-sample, one-step-ahead mean squared forecast error to be
$$
\msfein=\E\left[\frac{1}{T}\|\mathbf{Y}-\hat{\PhiB}\mathbf{Z}\|_2^2\ | \ \mathbf{Z}\right]=\tr(\mathbf{\Sigma}_u)+\frac1{T}\sum_{t=1}^T \left\|\sum_{\ell=1}^p(\hat\PhiB^{(\ell)} - \PhiB^{(\ell)})\mathbf{y}_{t-\ell}\right\|^2_2,
$$
with $\mathbf{Y}, \PhiB$ and $\mathbf{Z}$ as defined in equation \eqref{notation}.
While $\tr(\mathbf{\Sigma}_u)$ is the irreducible error, an unavoidable part of the forecast error, a good estimator of the autoregressive parameters should allow us to control the size of the second term. 
In Theorem \ref{mainresult}, we provide such a bound on the in-sample prediction error for the most flexible HLag method, namely elementwise HLag. 

\begin{thm} \label{mainresult}
	Suppose $T>\max\{25\log(pk^2),4\}$ and $pk^2\gg1$. Under Assumption \ref{assumptions} and taking all lag coefficients to be bounded in absolute value by $M$,  we choose \linebreak
	$\lambda \asymp v(\PhiB,\mathbf{\Sigma}_u)\sqrt{\log(pk^2)/T}$, where
	$v(\PhiB,\mathbf{\Sigma}_u) = \Lambda_{\text{max}}(\mathbf{\Sigma}_u)\left(1 + \frac{1+\mu_{\text{max}}(\PhiB)}{\mu_{\text{min}}(\PhiB)}\right)$.  Then, with probability at least $1-\frac{12}{(pk^2)^{23/2}}$,
	$$
	\frac1{T}\sum_{t=1}^T \left\|\sum_{\ell=1}^p(\hat\PhiB^{(\ell)} - \PhiB^{(\ell)})\mathbf{y}_{t-\ell}\right\|^2_2 \lesssim 
	M v(\PhiB, \mathbf{\Sigma}_u)\sqrt{\frac{\log(pk^2)}{T}}\sum_{i=1}^k\sum_{j=1}^kL_{ij}^{3/2},
	$$
	where $\hat\PhiB$ is the elementwise HLag estimator with $pmax = p$.
	\label{thm:lasso-slowrate}
\end{thm}
The proof of Theorem \ref{mainresult} is included in Section \ref{theory} of the appendix. Theorem \ref{mainresult} establishes in-sample prediction consistency in the high-dimensional regime $\log(pk^2)/T \rightarrow 0$. 
Hence, the same rate is obtained as for i.i.d.\ data, modulo a ``price" paid for dependence. The temporal and cross-sectional dependence affects the rate through the internal parameters $\Lambda_{\text{max}}(\mathbf{\Sigma}_u), \mu_{\text{min}}(\PhiB)$ and $\mu_{\text{max}}(\PhiB)$. 

While Theorem \ref{mainresult} is derived under the assumption that $p$ is the true order of the VAR, the results hold even if $p$ is replaced by any upper bound $pmax$ on the true order since the $\text{VAR}_k(p)$ can be viewed as a $\text{VAR}_k(pmax)$ with $\PhiB^{(\ell)} = {\bf 0}$ for $\ell > p$, see \cite{basu2013estimation}. The convergence rate then becomes $\sqrt{\log(pmax\cdot k^2)/T}$ instead of $\sqrt{\log(pk^2)/T}$.

The bound includes terms of the form $L_{ij}^{3/2}$.  The $3/2$ exponent can be removed if one adopts a more complicated weighting scheme (see e.g., \citealp{jenatton2011structured}, \citealp{bien2016convex}), which would avoid high order lag coefficients from being aggressively shrunken. However, in the context of VAR estimation, we find through simulation experiments that this aggressive shrinkage is in fact beneficial (see Section \ref{LOG} of the appendix).

%% ALGORITHM %%
\section{Optimization Algorithm} \label{Sec3}
We begin by noting that since the intercept $\boldsymbol{\nu}$ does not
appear in the penalty terms, it can be removed if we replace $\mathbf
Y$ by $\mathbf
Y({\bf I}_T-\frac1{T}{\bf11}^\top)$ and $\mathbf Z$ by $\mathbf Z({\bf
	I}_T-\frac1{T}{\bf11}^\top)$.
All three optimization problems are of the form 
\begin{align}
\min_{\PhiB}
\left\{ \frac{1}{2T}\|\mathbf{Y}-\PhiB\mathbf{Z}\|_2^2+\lambda\sum_{i=1}^k\sum_{\ell=1}^p\Omega_i(\PhiB^{(\ell:p)}_i)\right\},
\label{eq:hvar-general}
\end{align}
and \eqref{OV1}, \eqref{OV2}, and \eqref{OV3}
only differ by the form of the norm $\Omega_i$.
A key simplification is possible by observing that the objective above decouples across the rows of $\PhiB$:
$$
\min_{\PhiB}\sum_{i=1}^k\left[\frac1{2T}\|{\bf Y}_i-\PhiB_i{\bf Z}\|_2^2+\lambda\sum_{\ell=1}^p\Omega_i(\PhiB_i^{(\ell:p)})\right],
$$
in which $\mathbf{Y}_{i} \in\mathbb R^{1\times T}$ and $\PhiB_i=\PhiB_i^{(1:p)}\in\mathbb
R^{1\times kp}$.  
Hence, Equation \eqref{eq:hvar-general}
can be solved in parallel
by solving the ``one-row'' subproblem
%\begin{align}
%\label{HLassoAR}
$$\min_{\PhiB_i}\left\{\frac1{2T}\|{\bf Y}_i-\PhiB_i{\bf Z}\|_2^2+\lambda\sum_{\ell=1}^p\Omega_i(\PhiB_i^{(\ell:p)})\right\}.
$$
%\end{align}
\citet{jenatton} show that hierarchical group lasso problems can be
efficiently solved via the proximal gradient method.  This procedure can be viewed as an extension of traditional gradient descent
methods to nonsmooth objective functions.  Given a convex objective function
of the form $f_i(\PhiB_i)=\mathcal L_i(\PhiB_i)+\lambda \Omega_i^*(\PhiB_i)$, where $\mathcal L_i$ is differentiable
with a Lipschitz continuous gradient, the proximal gradient method
produces a sequence
$\hat\PhiB_i[1],\hat\PhiB_i[2],\ldots$ with the guarantee that 
$$f_i(\hat\PhiB_i[m])-\min_{\PhiB_i}f_i(\PhiB_i)$$
is $O(1/m)$ (cf. \citealt{beck}).  For $m=1,2,\ldots$, its update is given by
$$
\hat\PhiB_i[m]=\text{Prox}_{s_m\lambda\Omega^*_i}\left(\hat\PhiB_i[m-1]-s_{m}\nabla \mathcal L(\hat\PhiB_i[m-1])\right),
$$
where $s_m$ is an appropriately chosen step size and
$\text{Prox}_{s_m\lambda\Omega^*_i}$ is the proximal operator of the
function $s_m\lambda\Omega^*_i(\cdot)$, which is evaluated at the gradient
step we would take if we were minimizing $\mathcal L_i$ alone.  The proximal operator is defined as the unique solution of a
convex optimization problem involving $\Omega^*_i$ but not $\mathcal L_i$:
\begin{align}
\text{Prox}_{s_m\lambda\Omega^*_i}(u)=\argmin_{v}\left\{\frac1{2}\|u-v\|_2^2+s_m\lambda\Omega^*_i(v)\right\}.\label{eq:prox}
\end{align}
The proximal gradient method is particularly effective when the
proximal operator can be evaluated efficiently.  In our case, $\Omega_i^*(\PhiB_i)=\sum_{\ell=1}^p\Omega_i(\PhiB_i^{(\ell:p)})$ is a
sum of hierarchically nested Euclidean norms.  \citet{jenatton} show
that for such penalties, the proximal operator has
essentially a closed form solution, making it extremely efficient.
It remains to note that $\mathcal L_i(\PhiB_i)=\frac1{2T}\|\mathbf{Y}_{i}-\PhiB_{i}\mathbf{Z}\|^2_2$
has gradient $\nabla \mathcal L_i(\PhiB_i)=-\frac1{T}(\mathbf{Y}_{i}-\PhiB_{i}\mathbf{Z})\mathbf{Z}^\top$
and that the step size $s_m$ can be determined adaptively through a
backtracking procedure or it can be set to
the Lipschitz constant of $\nabla \mathcal
L_i(\PhiB_i)$, which in this case is $\sigma_1(\mathbf{Z})^{-2}$ (where $\sigma_1(\mathbf{Z})$ denotes the largest singular value of
$\mathbf{Z}$).

We use
an accelerated version of the proximal gradient
method 
which leads to a faster convergence rate and improved empirical
performance with minimal additional overhead.  
Our particular implementation is based on Algorithm 2 of
\citet{tseng2008accelerated}.  It repeats, for $m=1,2,\ldots$ to
convergence,
\begin{align*}
&\hat {\bm {\phi}}\leftarrow \hat\PhiB_{i}[m-1] + \theta_{m-1}(\theta_{m-2}^{-1} - 1)\left(\hat\PhiB_{i}[m-1]-\hat\PhiB_{i}[m-2]\right)\\
&\hat\PhiB_{i}[m] \leftarrow
\text{Prox}_{s_m\lambda\Omega^*_i}\left(\hat{\bm{\phi}}-s_m\nabla \mathcal
L_i(\hat{\bm{\phi}})\right), 
\end{align*}
with $\theta_m = 2/(m+2)$ as in \citet{tseng2008accelerated} and  converges at rate $1/m^2$ (compared to the unaccelerated proximal
gradient method's $1/m$ rate).
Alternatively, one could set $\theta_m = \frac{1}{2}\left(\sqrt{\theta_{m-1}^4 + 4\theta_{m-1}^2} - \theta_{m-1}^2\right)$ 
which is essentially the Fast Iterative Soft-Thresholding Algorithm developed by  \citet{beck}. We verified that our findings in the simulation study are unaffected by this choice.

Our full procedure is detailed in Algorithm \ref{alg1} and is applicable to
all three HLag estimators. 
Note that the algorithm requires an initial value $\hat\PhiB[0]$. 
As is standard in the regularization literature (e.g., \citealp{glmnet}), we use ``warm starts". We solve Algorithm \ref{alg1} for a grid of penalty values starting at $\lambda_{\text{max}}$, the smallest value of the regularization parameter
in which all coefficients will be zero. 
For each smaller value of $\lambda$ along this grid, we use the previous solution  as a ``warm start" $(\hat\PhiB[0])$ to run Algorithm \ref{alg1} with the new $\lambda$-value. A key advantage of our HLag estimates being solutions to a convex optimization problem is that the algorithms are stable and not sensitive to the choice of initialization \citep{beck}. 
As stopping criterion, we use $||\hat{\phi} - \hat{\boldsymbol\Phi}_i[m]||_\infty \leq \epsilon$, while one could also use $|| \hat{\boldsymbol\Phi}_i[m] - \hat{\boldsymbol\Phi}_i[m-1]||_\infty \leq \epsilon$. We opt for the former since we have numerically observed in our simulation experiments that considerably less iterations are needed without affecting  accuracy.

\begin{algorithm}[t]
	\caption{ \label{alg1} General algorithm for HLag with penalty $\Omega^*_i$}
	\begin{algorithmic} 
		\Require $\mathbf{Y},\mathbf{Z},\hat\PhiB[0],\lambda, \epsilon = 10^{-4}$  
		\State $\hat\PhiB[1]\leftarrow\hat\PhiB[0];\quad \hat\PhiB[2]\leftarrow\hat\PhiB[0]$
		\State $s\leftarrow \sigma_1(Z)^{-2}$
		\For{$i=1,\ldots,k$}
		\For{$m=3,4,\ldots$}
		\State $\hat {\bm{\phi}}\leftarrow \hat\PhiB_{i}[m-1]
		+\frac{m-2}{m+1}\left(\hat\PhiB_{i}[m-1]-\hat\PhiB_{i}[m-2]\right)$
		\State $\hat\PhiB_{i}[m]
		\leftarrow\text{Prox}_{s\lambda\Omega_i^*}\left(\hat{\bm{\phi} }+\frac{s}{T}\cdot(\mathbf{Y}_{i}-\hat{\bm{\phi}}\mathbf{Z})\mathbf{Z}^\top\right)$
		\If {$\|\hat{\bm{\phi}}-\hat\PhiB_{i}[m]\|_\infty\le\epsilon$} 
		\State \textbf{break}
		\EndIf
		\EndFor
		\EndFor\\
		\Return $\hat\PhiB[m]$
	\end{algorithmic}
\end{algorithm}

The algorithms for these methods differ only in the evaluation of their proximal operators (since each method has a different penalty $\Omega^*_i$).  However, all three choices of $\Omega^*_i$ correspond to hierarchical group lasso penalties,
allowing us to use the result of \citet{jenatton}, which shows that the proximal operator has a remarkably simple form.  We write these
three problems generically as 
\begin{align}
\hat\xx = \argmin_\xx\left\{\frac{1}{2}\|\xx-\tilde
\xx\|_2^2+\lambda\sum_{h=1}^Hw_h\|\xx_{g_h}\|_2\right\},\label{eq:generic-prox}
\end{align}
where $g_1\subset\cdots\subset g_H$.
The key observation in \citet{jenatton} is that the dual of the proximal problem \eqref{eq:prox} can be solved
exactly in a {\em single pass} of blockwise coordinate descent.  By
strong duality, this solution to the dual provides us with a solution
to problem \eqref{eq:prox}.
The updates of each block are extremely simple, corresponding to a groupwise-soft-thresholding
operation.  Algorithm \ref{alg2} shows the solution to \eqref{eq:generic-prox}, which includes all  three of our penalties as
special cases.

\begin{algorithm}[t]
	\caption{\label{alg2} Solving Problem \eqref{eq:generic-prox}}
	\begin{algorithmic}
		\Require $\tilde \xx, \lambda,w_1,\ldots, w_H$  
		\State $\r\leftarrow \tilde\xx$
		\For{$h=1,\ldots, H$}
		\State $\r_{g_h}\leftarrow (1 -\lambda w_{h}/\|\r_{g_h}\|_2)_+\r_{g_h}$
		\EndFor\\
		\Return $\r$ as the solution $\hat\xx$.
	\end{algorithmic}
\end{algorithm}

{\it Selection of the penalty parameters. \label{lambdaselection}}
While some theoretical results on the choice of penalty parameters are available in the literature \citep{basu2013estimation}, such theoretical results can not be used in practice since the 
penalty parameter's value  depends on properties of the underlying model that are not observable. For this reason, we use cross validation, one of the standard approaches to penalty parameter selection. 

Following \cite{Friedman10}, the grid of penalty values is constructed by starting
with $\lambda_{\text{max}}$, an estimate of the smallest value in which all coefficients are zero, then decrementing in log linear increments. The grid bounds are detailed in the appendix of \cite{Nicholson2017}. The HLag methods  rely on a single tuning parameter $\lambda$ in equation \eqref{eq:hvar-general}. Our penalty parameter search over a one-dimensional grid is  much less expensive than the search over a multi-dimensional grid as needed for the lag-weighted lasso \citep{BickelSong}. 
To accommodate the time series nature of our data, we select the penalty parameters using the cross-validation approach utilized by \cite{BickelSong} and \cite{BGR}.   
Given an evaluation period $[T_1,T_2]$, we use one-step-ahead {\em mean-squared
	forecast error} (MSFE) as a cross-validation score:
\begin{equation}
MSFE(T_1,T_2)=\frac{1}{k(T_2-T_1)}\sum_{i=1}^k \sum_{t=T_1}^{T_2-1}(\hat
{y}_{i,t+1}-{y}_{i,t+1})^2, \label{MSFEeq}
\end{equation}
with $\hat{y}_{i,t+1}$ representing the forecast for time
$t+1$ and component $i$ based on observing the series up to time $t$.  If multi-step ahead forecast horizons are desired, we can simply substitute \eqref{MSFEeq} with our desired forecast horizon $h$.
Since this penalty search requires looping over many time points, we have coded most of the HLag methods in \texttt{C++} to increase computational efficiency.

%% SIMULATIONS %%
\section{Simulation Study}
\label{Sec4}
We compare the proposed HLag methods with  13 competing approaches:
(i) AIC-VAR: least squares estimation of the VAR and selection of a universal lag order $\ell$ using AIC,
(ii) BIC-VAR: same as in (i) but lag order selection using BIC,
(iii) Lasso-VAR: estimation of the VAR using an $L_1$-penalty,
(iv) Lag-weighted (LW) Lasso-VAR: estimation of the VAR using a weighted $L_1$-penalty, which applies greater regularization to higher order lags,
(v) BGR-BVAR: Bayesian VAR of \cite{BGR},
(vi) GLP-BVAR: Bayesian VAR of \cite{GLP}, 
(vii) CCM-BVAR: Bayesian VAR of \cite{Carriero17}
(viii) DFM: Dynamic Factor Model (see e.g., \citealp{forni2000generalized}), 
(ix) FAVAR: Factor Augmented VAR \citep{Bernanke05}
(x) VAR(1): least squares estimation of a VAR(1)
(xi) AR: univariate autoregressive model,
(xii) Sample mean:  intercept-only model,
(xiii) Random walk:  vector random walk model.
The comparison methods are detailed in Section  \ref{benchmarks} of the appendix.

\subsection{Forecast Comparisons \label{simforecast}}
To demonstrate the efficacy of the HLag methods in
applications with various lag structures, we evaluate the proposed methods under four simulation scenarios.  

\begin{figure}[t]
	\begin{minipage}{0.5\linewidth}
		\begin{subfigure}{\textwidth}
			\centering
			\renewcommand{\thesubfigure}{1}
			\includegraphics[width=8cm]{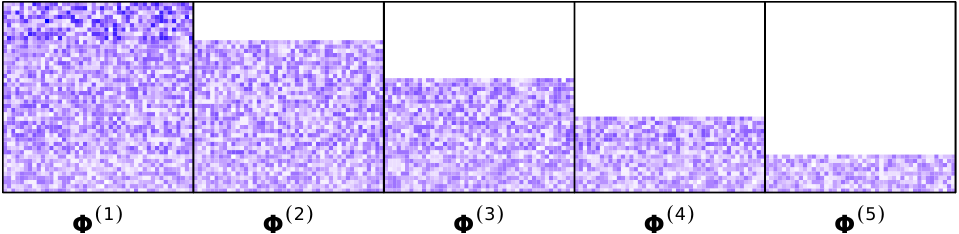}
			\caption{\em  Componentwise structure in Scenario 1.} 
		\end{subfigure}
	\end{minipage}
	\vspace{1cm}
	\begin{minipage}{0.5\linewidth}
		\begin{subfigure}{\textwidth}
			\centering
			\renewcommand{\thesubfigure}{2}
			\includegraphics[width=3.3cm]{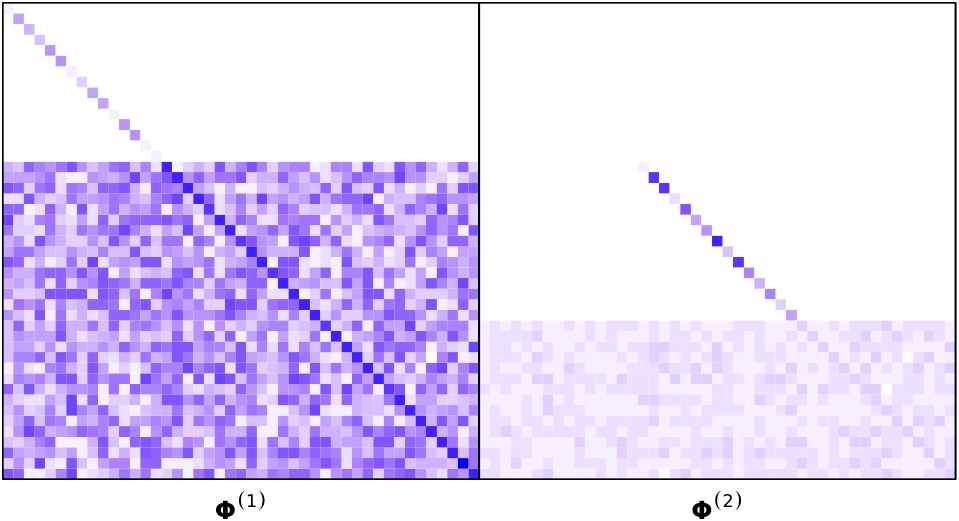}
			\caption{\em  Own-other structure in Scenario 2.} 
		\end{subfigure}
	\end{minipage}
	
	\vspace{-1cm}
	
	\begin{minipage}{0.5\linewidth}
		\begin{subfigure}{\textwidth}
			\centering
			\renewcommand{\thesubfigure}{3}
			\includegraphics[width=7.5cm]{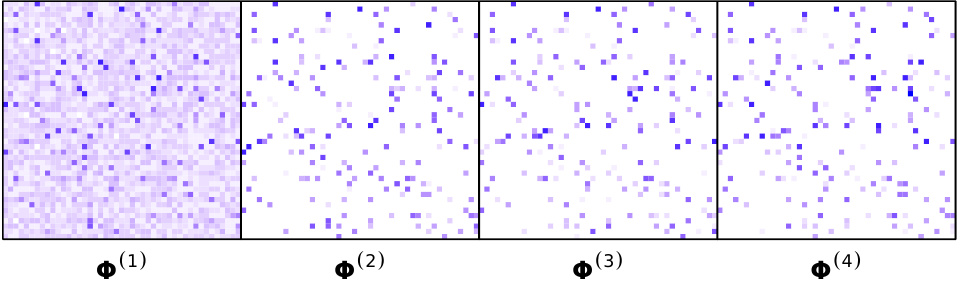}
			\caption{\em  Elementwise structure in Scenario 3.} 
		\end{subfigure}
	\end{minipage}
	\begin{minipage}{0.5\linewidth}
		\begin{subfigure}{\textwidth}
			\centering
			\renewcommand{\thesubfigure}{4}
			\includegraphics[width=7.5cm]{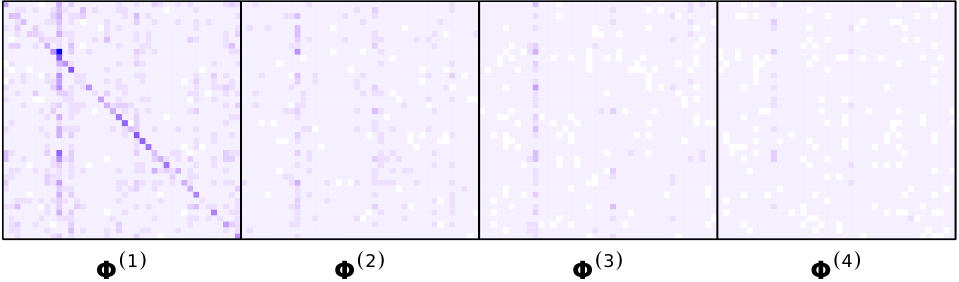}
			\caption{\em  Data-based structure in Scenario 4.}
		\end{subfigure}
	\end{minipage}
	\caption{Sparsity patterns (and magnitudes) of the HLag based simulation scenarios. Darker shading indicates coefficients that are larger in magnitude.}
	\label{figsim}
\end{figure}

In Scenarios 1-3, we take $k=45$ components, 
a series length of $T=100$ 
and simulate from a VAR with the respective HLag structures: 
componentwise,
own-other, and
elementwise.
In this section, we focus on simulation scenarios where the sample size $T$ is small to moderate compared to the number of parameters to be estimated $(pmax\cdot k^2 + k)$. We investigate the impact of increasing the time series length in Section \ref{increaseT} of the appendix. 
The coefficient matrices used in these scenarios  are depicted in Figure \ref{figsim}, panel (1)-(3) respectively.

In Scenario 4, we consider a data generating process (DGP) with $k=40$ and $T=195$ that does not a priori favor the HLag approaches vis-a-vis the
competing approaches but follows the ``data-based Monte Carlo method" \citep{ho1996} to make the simulation setting robust to arbitrary DGPs. 
This DGP does not have any special lag structure; all variables in all equations have $p=4$ non-zero lags, as can be seen from Figure \ref{figsim}, panel (4).

All simulations are generated from stationary coefficient matrices. Full details on each simulation design together with the steps taken to ensure the stationarity of the simulation structures are given in Sections
\ref{SimScenarios} and \ref{SimScenGen}
of the appendix.
In each scenario, the error
covariance is taken to be ${\bf\Sigma}_u=0.01\cdot\mathbf{I}_{k}$. 
We investigate the sensitivity of our results to various choices of error covariance in Section \ref{errorcov} of the appendix.
To reduce the influence of initial conditions on the DGPs, the first 500 observations were discarded as burn-in for each simulation run. We run $M=500$ simulations in each scenario. 

{\it Forecast performance measure.} 
We focus on the problem of obtaining reliable point forecasts.
To evaluate how well our methods and their competitors do in the context of providing such point forecasts, we measure 
their performance in terms of out-of-sample point forecast accuracy and choose mean squared forecast error as our main measure of performance.   
We generate time series of length $T$, fit the models to the first $T-1$ observations and use the last observation to compute the one-step-ahead \textit{mean squared forecast error}
\begin{equation}
MSFE= \dfrac{1}{kM} \sum_{s=1}^{M} \sum_{i=1}^k( {y}^{(s)}_{i, T}-\widehat{y}_{i,T}^{(s)})^2, \nonumber
\end{equation}
with
${y}_{i,T}^{(s)}$ the value of component time series $i$ at the time point $T$ in the $s^{th}$ simulation run, and  $\widehat{ y}_{i, T}^{(s)}$ is its predicted value. 

\begin{figure}
	\includegraphics[width=\textwidth]{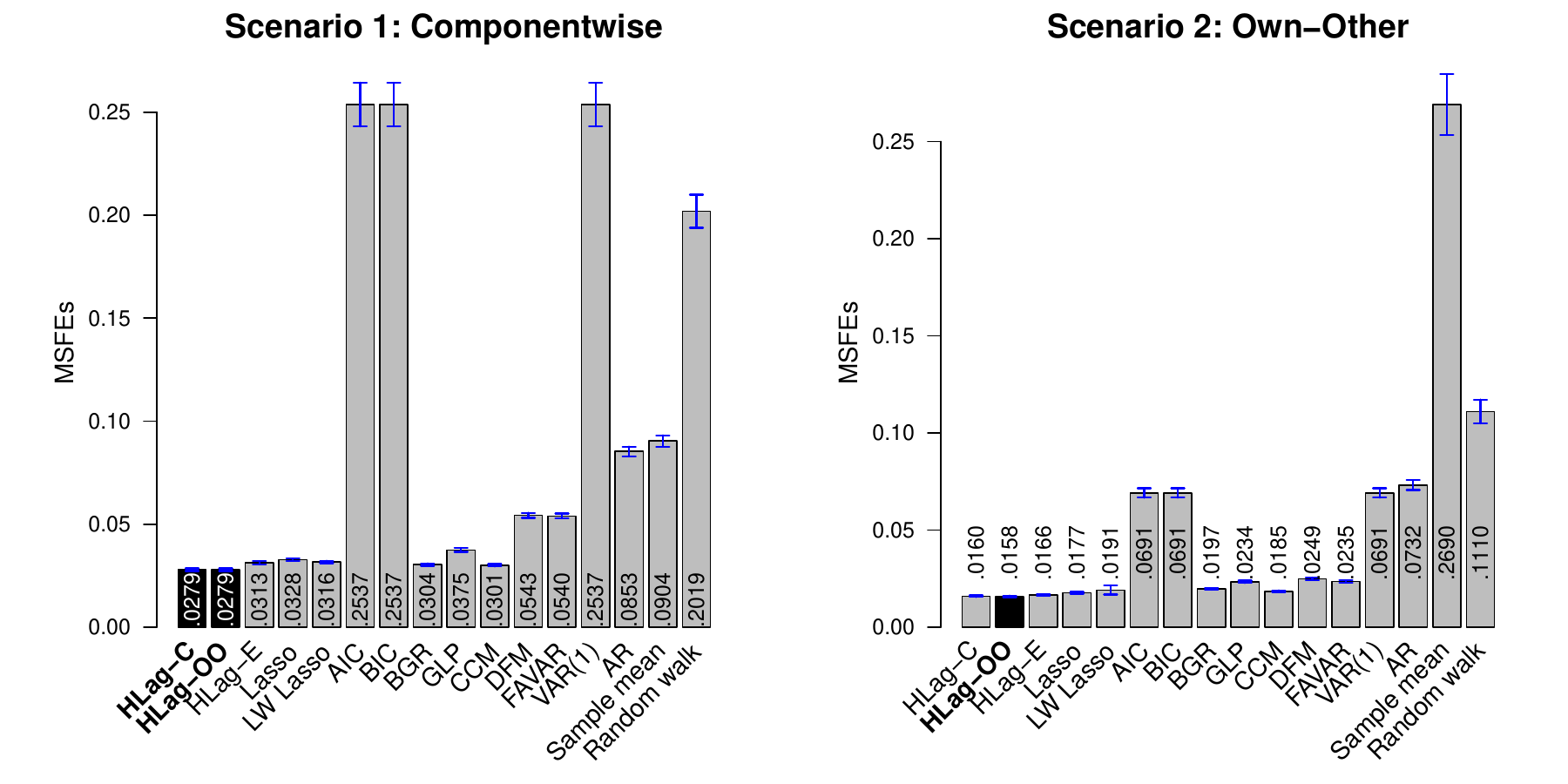}                
	\includegraphics[width=\textwidth]{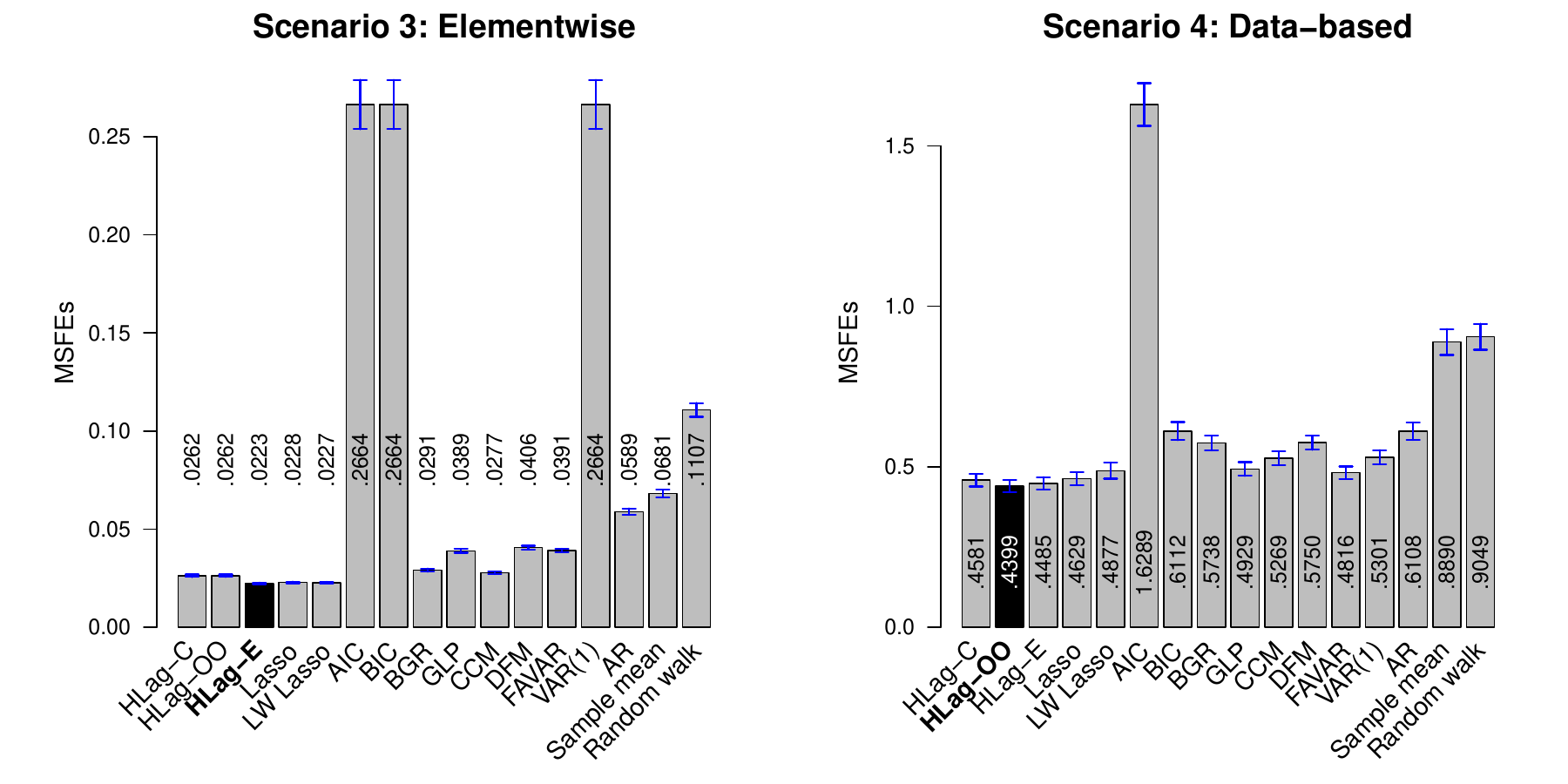} 
	\caption{Out-of-sample mean squared  forecast error  for VARs in Scenario 1 to 4. Error bars of length two standard errors are  in blue; the best performing method is  in black.\label{MSFEsim}}
\end{figure}

Figure \ref{MSFEsim} gives the forecast performance of the methods in Scenarios 1-4.
Concerning the VAR-based methods, we report the results for known ($p=5$ in Scenario 1, $p=2$ in Scenario 2 and $p=4$ in Scenario 3 and 4) maximal lag order.
We first discuss these results   and then summarize the differences in results when the maximal lag order is unknown, for which we take $pmax=12$.

{\it Scenario 1: Componentwise HLag.} Componentwise and own-other HLag perform best, which is to
be expected since both are geared explicitly toward Scenario 1's lag structure.  
Elementwise HLag outperforms the lag-weighted lasso, and both do better than the
lasso.  
Among the Bayesian methods, the BGR and CCM approaches are competitive to elementwise HLag, whereas the GLP approach is not. All Bayesian methods perform significantly worse (as confirmed with paired $t$-tests) than  componentwise and own-other HLag.
The factor models are not geared towards the DGP of Scenario 1: They select around five factors, on average, in their attempt to capture the time series dynamics and are not competitive to HLag. 
Regarding lag order selection with AIC/BIC, we can not estimate the VAR model for  $\ell>1$ with least squares, thus for a simple
benchmark we instead estimate a $VAR_k(1)$ by least squares.
Despite the explicit orientation toward modeling recent behavior in the $\text{VAR}_{45}(1)$ model, it
suffers both because it misses important longer range lag coefficients
and because it is an unregularized estimator of $\PhiB^{(1)}$ and
therefore has high variance.
The univariate AR benchmark also suffers because it misses the dynamics among the time series: its MSFE is more than twice as large as the MSFEs of the HLag methods. 

{\it Scenario 2: Own-other HLag.}
All three HLag methods perform  significantly  better than the competing methods.
As one would expect, own-other HLag achieves the best
forecasting performance, with componentwise and elementwise HLag performing only
slightly worse.  
As with the previous scenario, the least-squares approaches
are not competitive.

{\it Scenario 3: Elementwise HLag.} As expected, elementwise HLag outperforms all others.  
The lag-weighted lasso outperforms componentwise and own-other HLag, which is not surprising as it is designed to accommodate this type of structure in a more crude manner than elementwise HLag.  
The relatively poor performance of componentwise and own-other HLag is likely due to the coefficient matrix explicitly violating the structures in all 45 rows.  However, both still significantly outperform the  Bayesian methods,  factor-based methods and univariate benchmarks. 

{\it Scenario 4: Data-based.}
Though  all true parameters are non-zero, the HLag approaches perform considerably better than the lasso, lag-weighted lasso,  Bayesian,  factor-based and univariate approaches. 
HLag achieves variance reduction by enforcing sparsity and low max-lag orders. This, in turn, helps to improve forecast accuracy even for non-sparse DGPs where many of the coefficients are small in magnitude, as in Figure \ref{figsim}, panel (4). 

\begin{figure}
	\includegraphics[width=\textwidth]{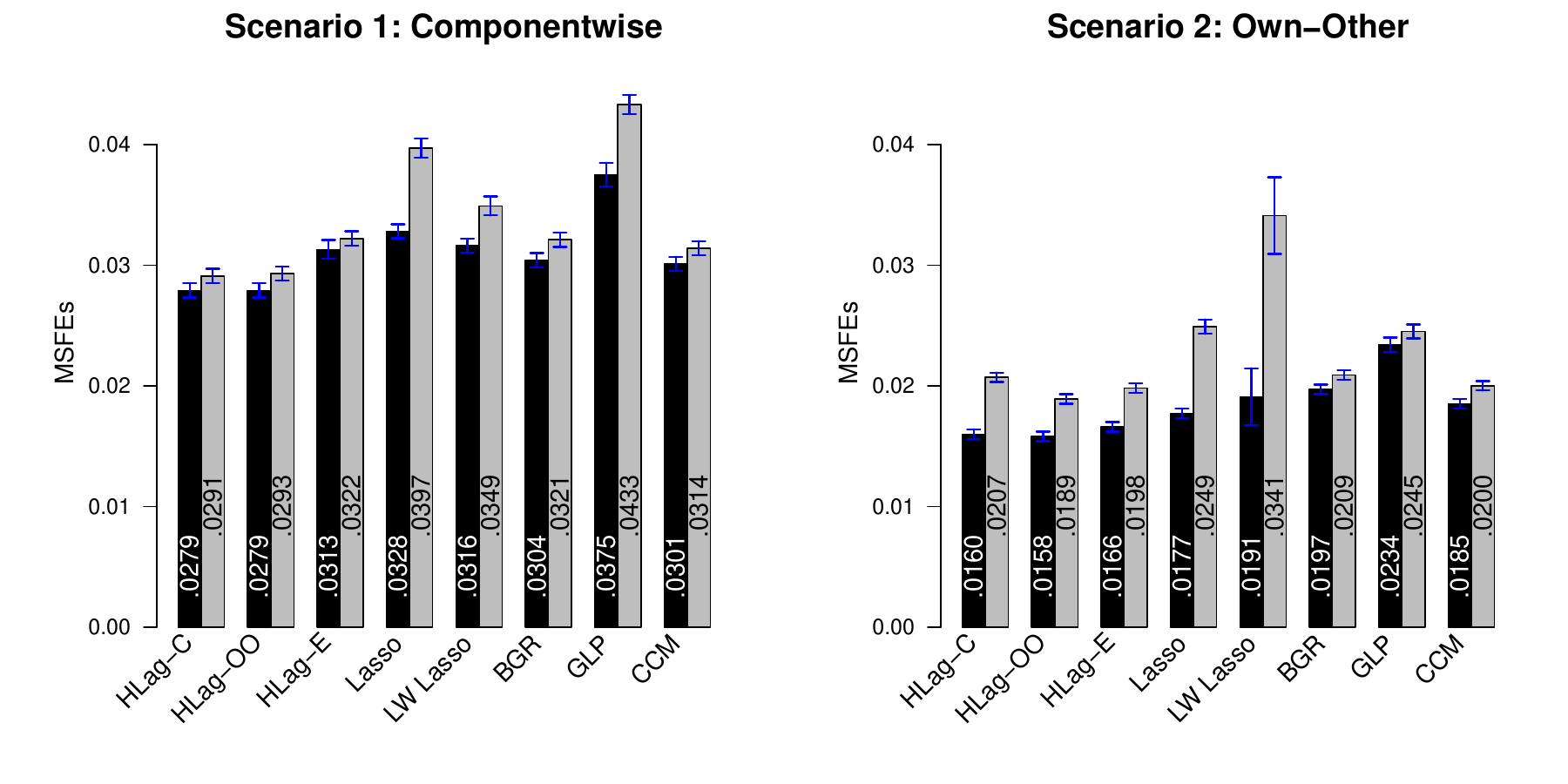}     
	\includegraphics[width=\textwidth]{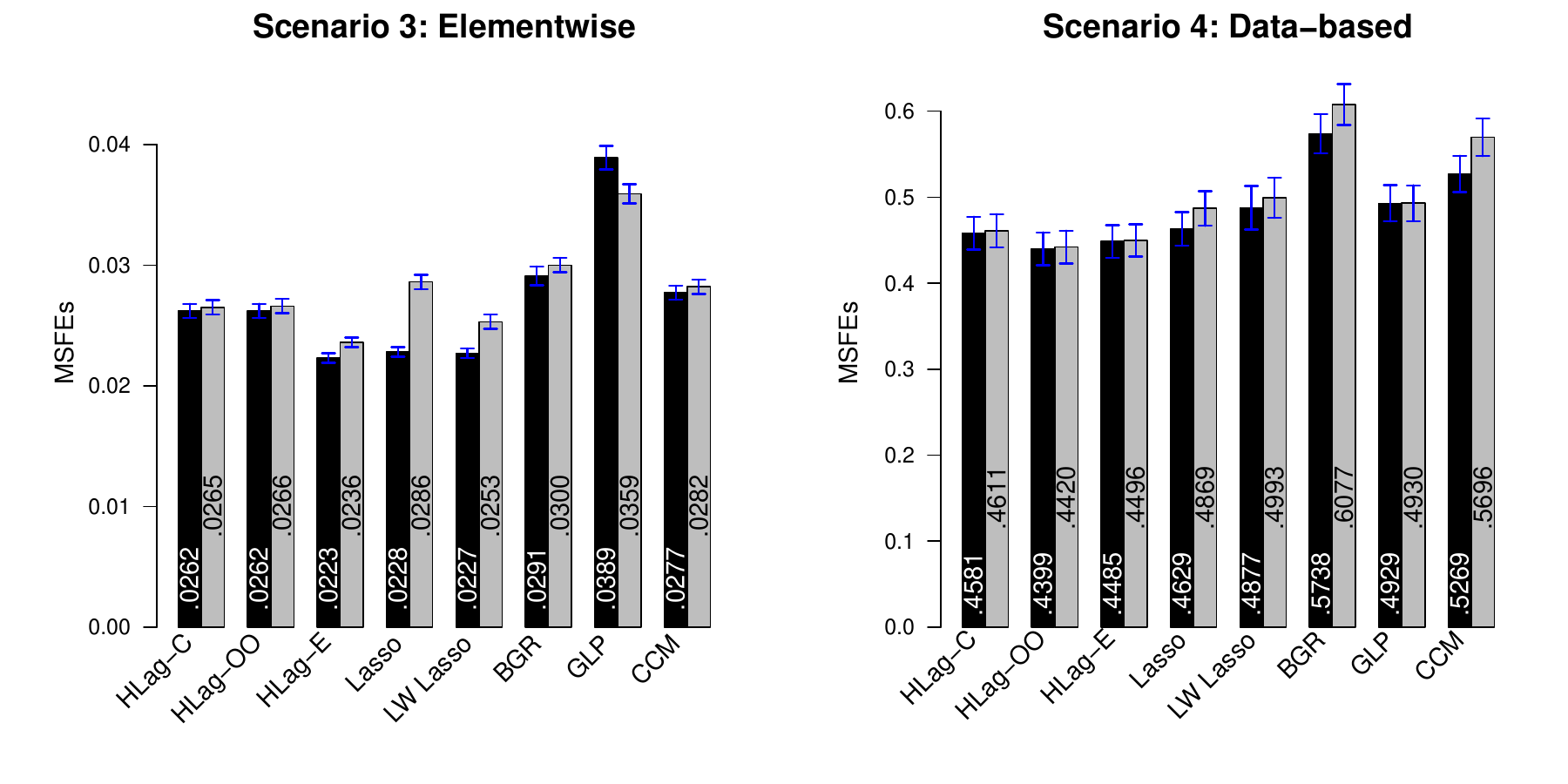} 
	\caption{Out-of-sample mean squared  forecast error  for VARs in Scenario 1 to 4 for known  (black) and unknown  (gray) order. Error bars of length two standard errors are  in blue.\label{MSFEs_log_scale_order_comparison}}
\end{figure}

{\it Unknown maximal lag order.}
In Figure \ref{MSFEs_log_scale_order_comparison}, we compare the performance of the VAR-based methods for known and unknown maximal lag order. For all methods in all considered scenarios, the MSFEs are, overall larger when the true maximal lag order is unknown since now the true lag order of each time series in each equation of the VAR can be overestimated. 
With a total of $pmax\cdot k^2 =12\times45^2$  autoregressive parameters to
estimate, the methods that assume an ordering, like HLag, are  greatly advantaged over a
method like the lasso that does not exploit this knowledge.
Indeed, in Scenario 3 with unknown order, componentwise and own-other HLag outperform the lasso. 

{\it Computation time.}
Average computation times, in seconds on an Intel
Core i7-6820HQ  2.70GHz machine including the penalty parameter search, for Scenario 1 and known order are reported in Table \ref{computation} for comparison. 
The relative performance of the methods with regard to average computation time in the other scenarios was very similar. 
The HLag methods have a clear advantage over the Bayesian methods of \cite{GLP}, \cite{Carriero17} and the lag-weighted lasso. The latter minimally requires specifying a weight function, and a two-dimensional penalty parameter search in our implementation, which is much more time intensive than a one-dimensional search, as required for HLag.  The Bayesian method of \cite{BGR} is fast to compute since there is a closed-form expression for the mean of the posterior distribution of the autoregressive parameters conditional on the error variance-covariance matrix. 
While the  Bayesian method of \cite{BGR} and lasso require, in general, less computation time, HLag has clear advantages over the former two in terms of  forecast accuracy, especially when the maximal lag length $pmax$ is large, but also in terms of lag order selection, as discussed in the following sections.

\begin{table}[t]
	\centering
	\begin{tabular}{llS} \hline 
		Class 		& Method & Computation \ time \ (in \ seconds)\\ \hline 
		HLag 		& Componentwise     & 17.1\\
		& Own-other         & 6.5\\
		& Elementwise       & 10.9\\
		VAR     	& Lasso             & 8.4 \\
		& Lag-weighted lasso & 154.2\\
		BVAR     	& BGR               & 0.4\\
		& GLP & 348.8 \\ 
		& CCM & 79.5\\
		Factor 		& DFM & 3.5 \\│
		& FAVAR & 3.1 \\ \hline 
	\end{tabular}
	\caption{Average computation times (in seconds), including the penalty parameter search, for the different methods in Scenario 1 ($T=100, \ k=45, \ p=5$). The results for the least squares, sample mean, VAR(1), AR model and random walk are omitted as their computation time is negligible. \label{computation}}
\end{table}

\subsection{Robustness of HLag as $pmax$ Increases \label{robustness}}
We  examine the impact of the maximal lag order $pmax$ on HLag's performance.  
Ideally, provided that $pmax$ is large enough to capture the system dynamics, its choice should have little impact on forecast performance.  However, we  expect regularizers that treat each coefficient democratically, like the lasso, to experience degraded forecast performance as $pmax$ increases.  

\begin{figure}
	\centering
	\includegraphics[width=8.5cm]{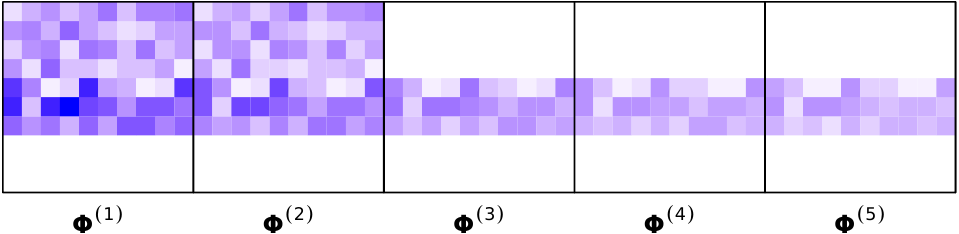}
	\caption{Componentwise  structure in the Robustness simulation Scenario 5. \label{fig:figrob}}
\end{figure}%

\begin{figure}
	\centering	
	\includegraphics[width=6.9cm]{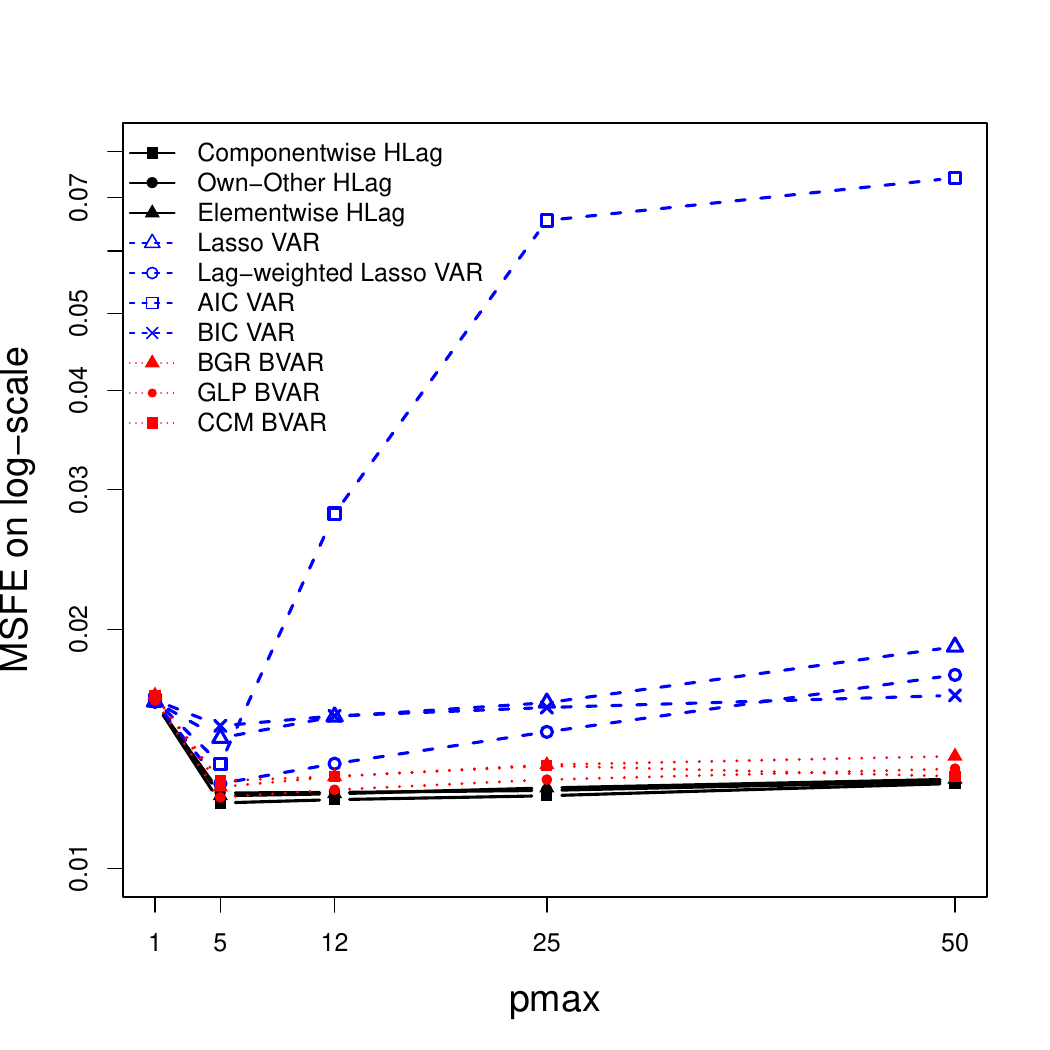} 	
	\caption{Robustness simulation scenario: Out-of-sample  mean squared  forecast errors, for different values of the maximal lag order $pmax$.}\label{Sim4MSFE}
\end{figure}

As an experiment, we simulate from an $\text{HLag}_{10}^{C}(5)$ while increasing $pmax$ to substantially exceed the true $\mathbf{L}$.  Figure \ref{fig:figrob} depicts the coefficient matrices and its magnitudes in what we will call Scenario 5. All series in the first 4 rows have $\mathbf{L}=2$, the next 3 rows have $\mathbf{L}=5$, and the final 3 rows have $\mathbf{L}=0$.  
We consider varying $pmax\in \{1,5,12,25,50\}$ and show the MSFEs of all VAR-based methods requiring a maximal lag order in Figure \ref{Sim4MSFE}.  As $pmax$ increases, we  expect the performance of HLag to remain relatively constant whereas the lasso and information-criterion based methods should return worse forecasts.  

At $pmax=1$ all models are misspecified.  Since no method is capable of capturing the true dynamics of series 1-7 in Figure \ref{fig:figrob}, all perform poorly.  
As expected, after ignoring $pmax=1$, componentwise HLag achieves the best performance across all other choices for $pmax$, but is very closely followed by the own-other and elementwise HLag methods.  
Among the information-criterion based methods, AIC performs substantially worse than BIC as $pmax$ increases.  
This is likely the result of BIC assigning a larger penalty on the number of coefficients than AIC.  
The lasso's performance degrades substantially as the lag order increases, while the lag-weighted lasso and Bayesian methods are somewhat more robust to the lag order, but still achieve worse forecasts than every HLag procedure under all choices for $pmax$.  

\subsection{Lag Order Selection}
While our primary intent in introducing the HLag framework is better point
forecast performance and improved interpretability, one can also view
HLag as an approach for selecting lag order.  Below, we examine  the performance of the proposed methods in estimating the maxlag matrix
$\LL$ defined in Section \ref{sec:lag-struct-hier}.
Based on an estimate $\hat\PhiB$ of the autoregressive coefficients, we
can likewise define a matrix of estimated lag orders:
$$
\hat{\LL}_{ij}=\max\{\ell:\hat{\PhiB}_{ij}^{(\ell)}\neq0\},
$$
where we define $\hat\LL_{ij}=0$ if $\hat\PhiB_{ij}^{(\ell)}=0$ for all $\ell$.
It is well known in the regularized regression literature (cf., \citealt{leng2006note}) that the optimal tuning
parameter for prediction is different from that for support recovery.
Nonetheless, in this section we will proceed with the
cross-validation procedure used previously with only two minor modifications intended to
ameliorate the tendency of cross-validation to select a value of
$\lambda$ that is smaller than optimal for support recovery.
First, we cross-validate a relaxed version of the regularized methods
in which the estimated nonzero coefficients are refit using ridge
regression, as detailed in Section \ref{RVARappendix} of the appendix.  
This modification makes the MSFE more sensitive to
$\hat{\LL}_{ij}$ being larger than necessary.  Second, we use the ``one-standard-error rule''
discussed in \cite{elements}, in which we select the largest value of
$\lambda$ whose MSFE is no more than one standard error above that of
the best performing model (since we favor the most parsimonious model
that does approximately as well as any other).  

\begin{figure}
	\centering 
	\includegraphics[width=\textwidth]{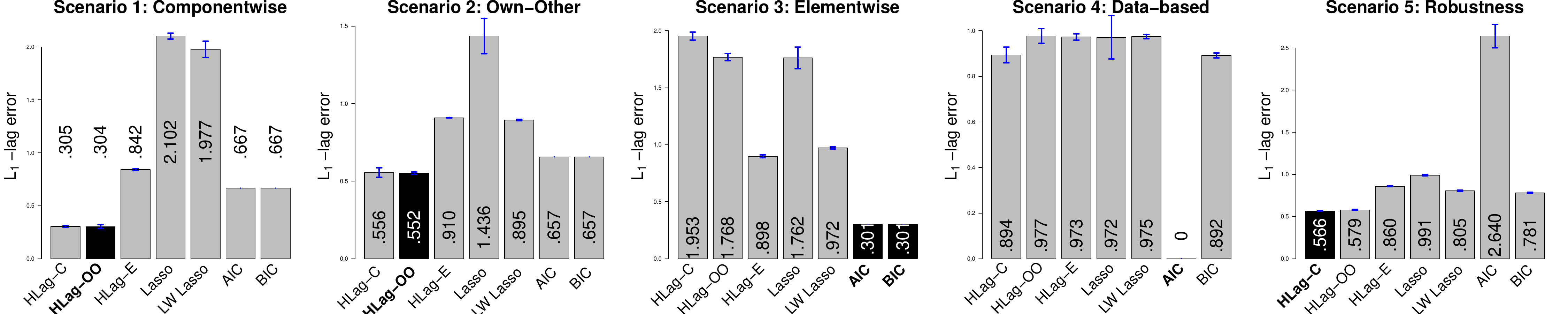}                 
	\caption{$L_1$-lag selection performance for Scenario 1 to 5. Error bars of length two standard errors are  in blue; the best performing method is  in black. \label{lagsim}}
\end{figure}

We consider Scenario 1 to 5  and estimate a $VAR_k(12)$. A procedure's lag order selection accuracy is measured based on the
sum of absolute differences between $\LL$
and $\hat{\LL}$  and the maximum absolute differences between  $\LL$
and $\hat{\LL}$:
\begin{equation}
\|\hat\LL-\LL\|_1=\sum_{ij} |\hat{\LL}_{ij} - \LL_{ij}| \  \text{and} \ \|\hat\LL-\LL\|_\infty=\max_{i,j} |\hat{\LL}_{ij} - \LL_{ij}|. \label{Laccuracy}
\end{equation}
The former can be seen as an overall measure of lag order error,  the latter  as a ``worst-case" measure. 
We present the  values on both measures  relative to that of
the sample mean (which chooses $\hat\LL_{ij} = 0$ for all i and j).
Figure \ref {lagsim} gives the results on the $L_1$-based measure.
We focus our discussion on the VAR-methods  performing actual lag order selection.
We first discuss these results  then summarize the differences in results for the $L_\infty$-based measure. 

{\it $L_1$-lag selection performance.}
In Scenarios 1-3, the HLag methods geared towards the design-specific lag structure perform best, as expected. 
Least squares AIC/BIC always estimates a $VAR_k(1)$ and performs considerably worse than the best performing HLag method in Scenarios 1-2.  
In Scenario 3, 
they attain the best performance since around 82\% of the elements in the true maxlag matrix are equal to one, and hence correctly recovered. However,  the higher order dynamics of the remaining 18\% of the elements are ignored, while elementwise HLag---which performs second best---better captures these dynamics. This explains why in terms of MSFE, elementwise HLag outperforms the $VAR_k(1)$ by a factor of 10.

In Scenario 4,  least squares AIC consistently recovers the true universal order $p=4$. Nevertheless, it has, in general, a tendency to select the highest feasible order, which  happens to coincide here with the true order. Its overfitting tendency generally has more negative repercussions, as can be seen from  Scenario 5, and even more importantly from its poor forecast performance.
Componentwise HLag and  least squares BIC perform similarly and are second best. Own-other, elementwise HLag, lasso and lag-weighted lasso perform similarly but underestimate the lag order of the component series with small non-zero values at higher order lags. While this negatively affects their lag order selection performance, it helps for forecast performance as discussed in  Section \ref{simforecast}.

In Scenario 5, 
componentwise and own-other HLag achieve the best performance. 
Their performance is five times better than the least squares AIC, and roughly 1.5 times better than the lasso, lag-weighted lasso and least squares BIC. 
Elementwise HLag substantially outperforms the lasso and least squares AIC, which consistently severely overestimates the true lag order. The  least squares BIC, on the other hand, performs similarly to elementwise HLag on the lag selection criterion but selects the universal lag order at either 1 or 2 and thus does not capture the true dynamics of series 5-7 in Figure \ref{fig:figrob}.

\begin{figure}
	\centering	
	\includegraphics[width=6.3cm]{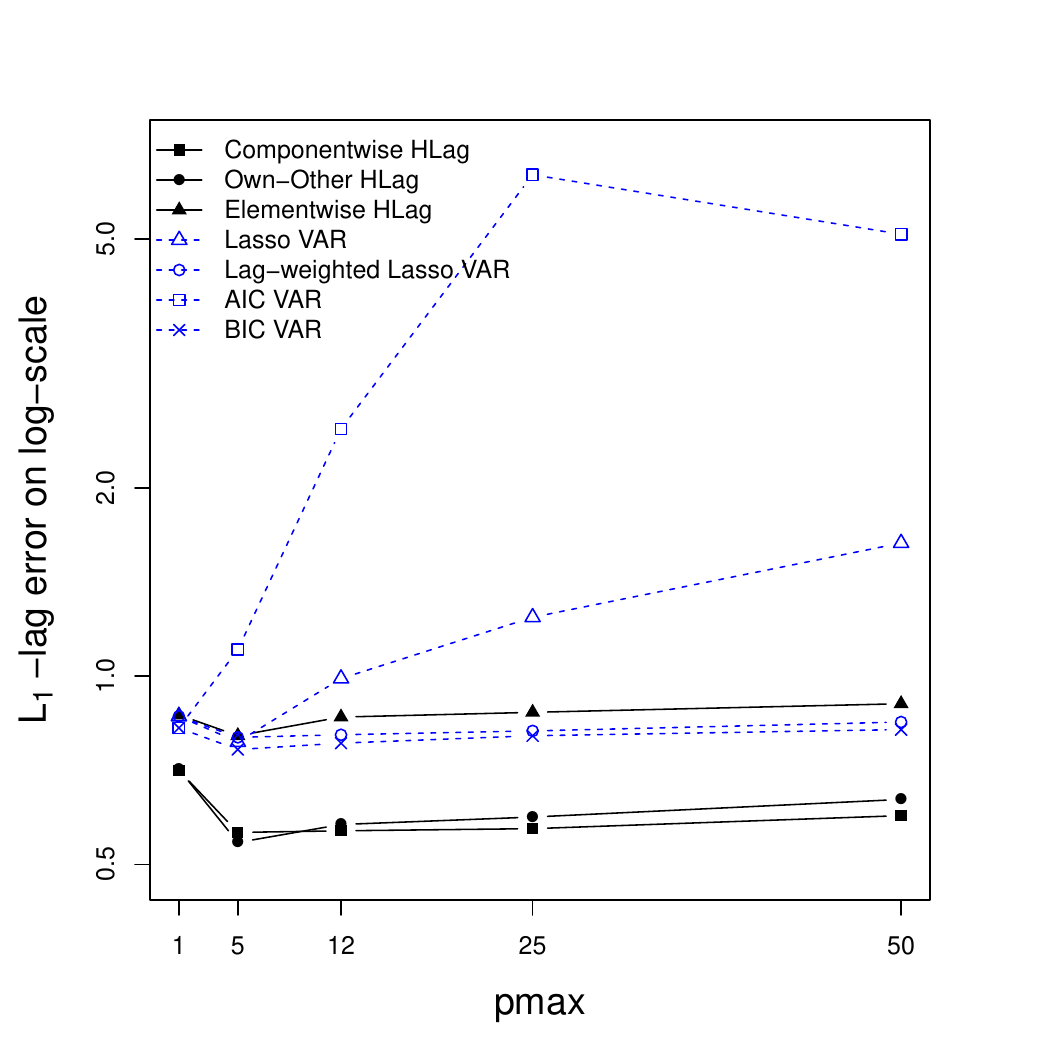}
	\includegraphics[width=6.3cm]{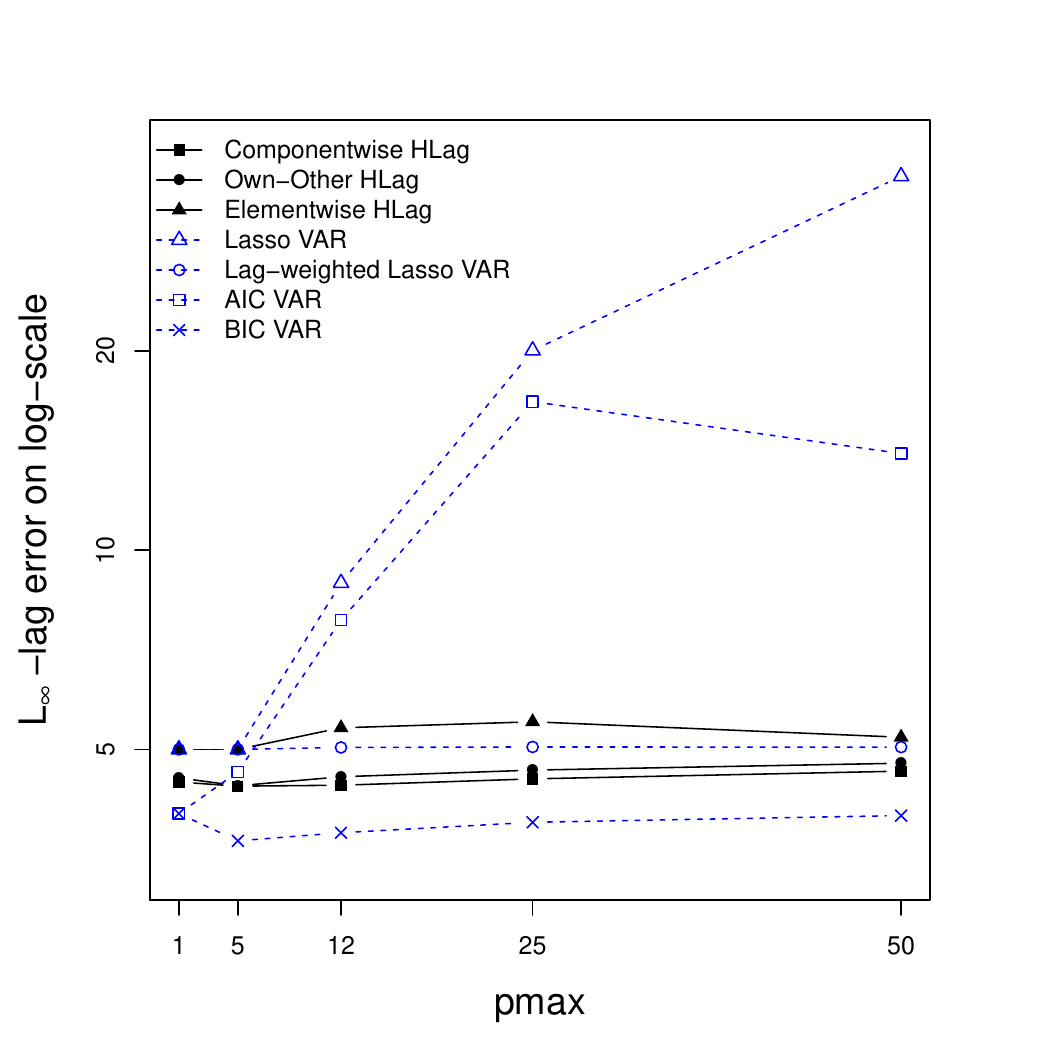}
	\caption{Robustness simulation scenario: Lag order error measures, for different values of the maximal lag order $pmax$.}\label{Sim4lags}
\end{figure}

In Figure \ref{Sim4lags}, we examine  the impact of the maximal lag order $pmax$ on a method's lag order error. At the true order ($pmax = 5$), all methods achieve their best performance. As $pmax$ increases, we find the methods' performance to decrease, in line with the findings by \cite{Percival12}. Yet, the HLag methods and lag-weighted lasso remain much more robust than the AIC and lasso, whose performance degrade considerably.

\begin{figure}
	\centering  
	\includegraphics[width=\textwidth]{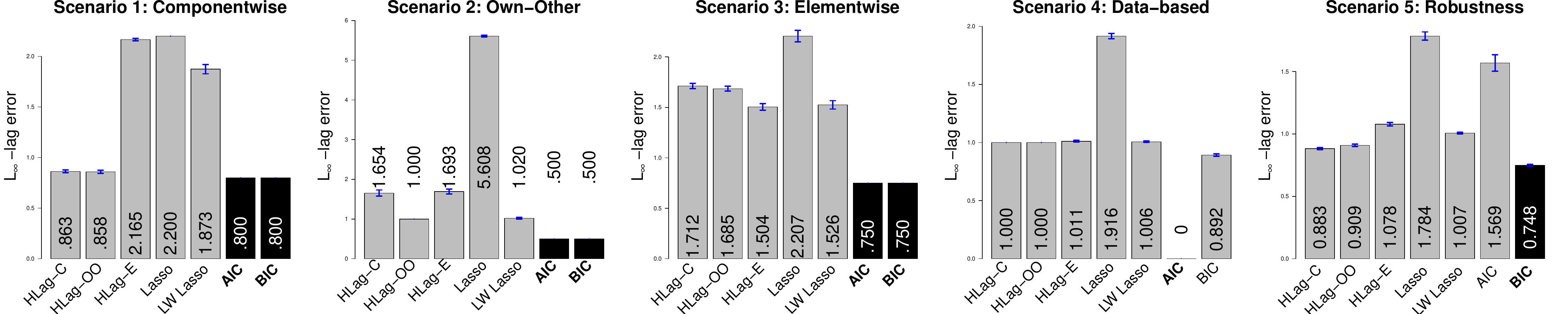}                
	\caption{$L_\infty$-lag selection performance for Scenario 1 to 5. Error bars of length two standard errors are  in blue; the best performing method is  in black.\label{lagsimLinf}}
\end{figure}

{\it $L_\infty$-lag selection performance.}
Results on the ``worst-case" $L_\infty$-measure are presented in Figure \ref{lagsimLinf}. 
Differences compared to the $L_1$-measure are:
(i)  Least squares AIC/BIC are the best performing. This occurs since the true maximal lag orders are small, as well as the estimated lag orders by AIC/BIC due to the maximum number of parameters that least squares can take.  Hence, the maximal difference between both is, overall,  small. Their negative repercussions are better reflected through the overall $L_1$-measure, or in case of the AIC as $pmax$ increases (see Figure \ref{Sim4lags}).
(ii) Componentwise  and own-other HLag are more robust with respect to the $L_\infty$-measure than elementwise HLag. The former two either add an additional lag for all time series or for none, thereby encouraging low lag order solutions---and thus controlling the maximum difference with the small true orders---even more than elementwise HLag. The latter (and the lag-weighted lasso) can flexibly add an additional lag  for each time series separately. Their price to pay for this flexibility becomes apparent through the $L_\infty$-measure.
(iii) A noticeable difference occurs between the methods that assume an ordering, like HLag and the lag-weighted lasso, and methods, like the lasso, that do not encourage low maximal lag orders. 
The lasso often picks up at least one lag close to the maximally specified order, thereby explaining its bad performance in terms of the $L_\infty$-measure. As $pmax$ increases, its performance deteriorates even more, see Figure \ref{Sim4lags}.

{\it Stability across time.} We  verified the stability in lag order selection across time  with a rolling window approach. We estimate the different models for the last 40  time points (20\%), each time using the most recent 160 observations. For each of these time points, the lag matrices are obtained and the lag selection accuracy measures in equation \eqref{Laccuracy} are computed. For all methods, we find the lag order selection to be very stable across time with no changes in their relative performance.

%% DATA %%
\section{Data Analysis}\label{Sec5}
We  demonstrate the usefulness of the proposed HLag  methods for various applications.
Our first and main  application is macroeconomic forecasting (Section \ref{macro}). We investigate the performance of the HLag methods on several VAR models where the number of time series is varied relative to the fixed sample size.
Secondly, we use the HLag methods for forecast applications with high sampling rates (Section \ref{highsampling}).

For all applications, we compare the forecast performance of the HLag methods to their competitors.
We use the cross-validation approach from Section \ref{lambdaselection} for  penalty parameter selection on time points $T_1$ to $T_2$:
At each time point $t=T_1-h, \ldots, T_2-h$  (with $h$ the forecast horizon), we first standardize each series to have sample mean zero and variance one using the most recent $T_1-h$ observations. We do this to account for possible time variation in the first and second moment of the data. Then, we estimate the VAR with $pmax$ and compute the weighted Mean Squared Forecast Error 
\begin{equation}
wMSFE =\frac{1}{k(T_2 - T_1 + 1)} \sum_{i=1}^{k} \sum_{t=T_1-h}^{T_2 -h} \left(\frac{y_{i, t+h}^{(s)} - \hat{y}_{i, t+h}^{(s)}}{\widehat{\sigma}_i}\right)^2, \nonumber \label{wMSFE}
\end{equation}
where $\widehat{\sigma}_i$ is the standard deviation of the $i^{th}$ to be forecast series, computed over the forecast evaluation period $[T_1,T_2]$ for each penalty parameter. We use a \textit{weighted} MSFE to account for the different volatilities and predictabilities of the different series when computing an overall forecast error measure \citep{carriero11}. The selected penalty parameter is  the one giving the lowest $wMSFE$.

After penalty parameter selection, time points  $T_3$ to $T_4$ are used for out-of-sample rolling window forecast comparisons.  Again, we standardize each series separately in each rolling window, estimate a VAR on the most recent $T_3-h$ observations and evaluate the overall forecast accuracy with the $wMSFE$ of equation \eqref{wMSFE}, averaged over all $k$ time series and  time points of the forecast evaluation period. Similar results are obtained with an expanding window forecast exercise and  available from the authors upon request. 

Finally, to assess the statistical significance of the results, we use the Model Confidence Set (MCS) procedure of \cite{Hansen11}. 
It separates the
best forecast methods with equal predictive ability from the others, who perform significantly worse. We use the \verb|MCSprocedure| function in \verb|R| to obtain a MCS that contains the best model with 75\% confidence as done in \cite{Hansen11}. 

\subsection{Macroeconomic Forecasting} \label{macro}
We apply the proposed HLag  methods to a collection of US macroeconomic time series compiled by \cite{stockdataset} and augmented by \cite{koop}.  
The full data set, publicly available at The Journal of Applied Econometrics Data Archive,  contains 168 quarterly macroeconomic indicators 
over 45 years: Quarter 2, 1959 to  Quarter 4, 2007, hence $T=195$.
Following \cite{stock12}, we classify the series into 13 categories, listed in Table \ref{SWgroups} of the appendix.
Further details can be found in Section \ref{SWAppendix} of the appendix.

Following \cite{koop}, we estimate four VAR models on this data set:
The \emph{Small-Medium VAR ($k=10$)} which consists of GDP growth rate, the Federal Funds Rate, and CPI plus 7 additional variables, including monetary variables. 
The \textit{Medium} VAR ($k=20$) which contains the Small-Medium group plus 10 additional variables containing aggregated information on several aspects of the economy. 
The \emph{Medium-Large VAR ($k=40$)} which contains the Medium group plus 20 additional variables, including most of the remaining aggregate variables in the data set.
The \emph{Large VAR ($k=168$)} which contains the Medium-Large group plus 128 additional variables, consisting primarily of the components that make up the aggregated variables.
Note that the number of parameters quickly increases from $4\times10^2+10 = 410$ (\textit{Small-Medium} VAR) over $4\times20^2+20 = 1,\!620$ (\textit{Medium} VAR), $4\times40^2+40 = 6,\!440$ (\textit{Medium-Large} VAR), to $4\times168^2+168 = 113,\!064$ (\textit{Large} VAR). 

\subsubsection{Forecast Comparisons} \label{forecast_comparison}
We compare the forecast performance of the HLag methods to their competitors on the four VAR models with $pmax=4$, following the convention from \cite{koop}.
Quarter 3, 1977 ($T_1$) to Quarter 3, 1992 ($T_2$) is used for penalty parameter selection; Quarter 4, 1992 ($T_3$) to Quarter 4, 2007 ($T_4$) are used for out-of-sample rolling window forecast comparisons.
We start with a discussion on the forecast accuracy for all series combined, then break down the results 
across different VAR sizes for specific variables.

\begin{figure} 
	\includegraphics[width=\textwidth]{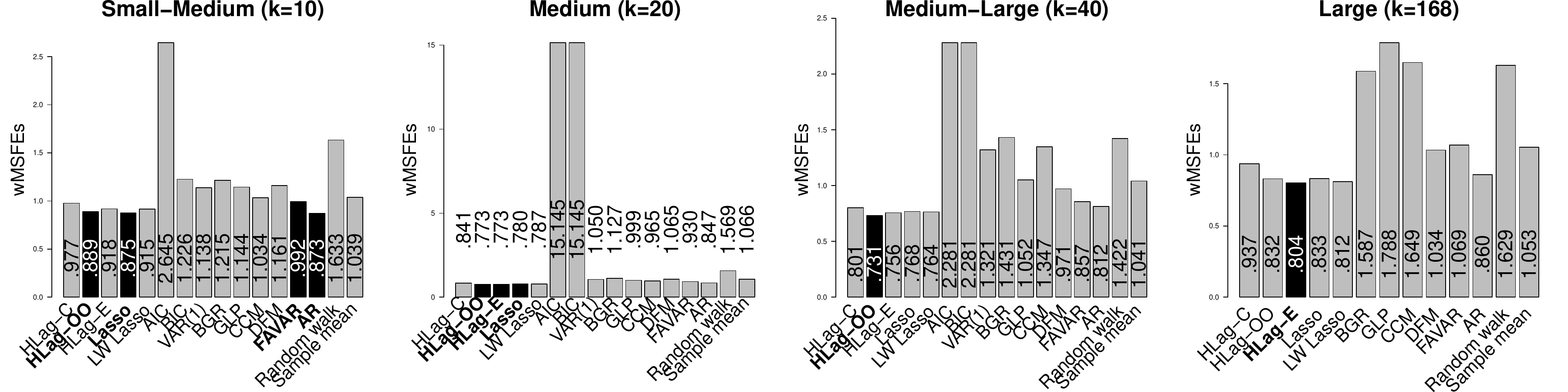} 
	\caption{\label{h1wMSFEall} Rolling out-of-sample one-step-ahead $wMSFE$
		for the four VAR sizes. For each VAR size, forecast methods in the 75\% Model Confidence Set (MCS) are in black.}
\end{figure}

{\it Forecast performance across all series.}
We report the out-of-sample one-step-ahead weighted mean squared forecast errors for the four VAR groups with forecast horizon $h=1$ in Figure \ref{h1wMSFEall}.  
We discuss the results for each VAR group separately since the $wMSFE$ are not directly comparable across the panels of Figure \ref{h1wMSFEall}, as an average is taken over different component series which might be more or less difficult to predict. 

With only a limited number of component series $k$ included in the \textit{Small} VAR, the univariate AR attains the  lowest $wMSFE$, but own-other HLag, the lasso and FAVAR have equal predictive ability since they are included in the MCS.
As more component series are added  in the \textit{Medium} and \textit{Medium-Large} VAR,   own-other and elementwise HLag outperform all other methods. 
The more flexible own-other and elementwise structures perform similarly, and better than the componentwise structure.
While the MCS includes own-other HLag, elementwise HLag and the lasso for the \textit{Medium} VAR,  only own-other HLag survives for the \textit{Medium-Large} VAR. 
This supports the widely held belief that in economic applications, a components' \emph{own} lags are likely more informative than \emph{other} lags and that maxlag varies across components. 
Furthermore, the Bayesian and factor models are never included in the MCS, nor are the least squares methods, or univariate methods. For the \textit{Medium} VAR, the information criteria AIC and BIC always select three lags. Since a relatively large number of parameters need to be estimated, their estimation error becomes large, and this, in turn, severely impacts their forecast accuracy. 

Next, consider the \emph{Large} VAR, noting that the VAR by AIC, BIC and VAR(1) are overparametrized and not included. As the number of component series $k$ further increases, the componentwise HLag structure becomes less realistic.  
This is especially true in high-dimensional economic applications, in which a core subset of the included series is typically most important in forecasting.
In Figure \ref{h1wMSFEall} we indeed see that the more flexible own-other and elementwise HLag perform considerably better than the componentwise HLag.  
The MCS confirms the strong performance of elementwise  HLag.

HLag's good performance across all series is confirmed by forecast accuracy results broken down by macroeconomic category. The flexible elementwise HLag is the best performing method; for almost all categories, it is included in the MCS, which is not the case for any other  forecasting method. Detailed results can be found in Figure \ref{wMSFEcategories} of the appendix. 

Furthermore, our findings remain stable when we increase the maximal lag order $pmax$. In line with \cite{BGR}, we re-estimated all models with $pmax = 13$. Detailed results are reported in Figure \ref{h1wMSFEallpmax13} of the appendix.
For the \textit{Small-Medium} VAR, own-other HLag performs comparable to the AR benchmark, while it  outperforms all other methods for larger VARs.
The lasso (and to a lesser extent the lag-weighted lasso) loses its competitiveness vis-a-vis the HLag approaches as soon as   the maximal lag order $pmax$ increases, in line with the results of Section \ref{robustness}. 

Finally, we re-did our forecast exercise for longer forecast horizons $h=4$ and $h=8$. Detailed results are reported in Figure \ref{h4-8-wMSFEall} of the appendix. 
All forecast errors increase with distant forecast horizons.
Nonetheless, own-other HLag remains among the best forecast methods: it is the only method that is always included in the MCS. Its performance gets closer to the sample mean as the forecast horizon increases.

\begin{figure}    
	\includegraphics[width=\textwidth]{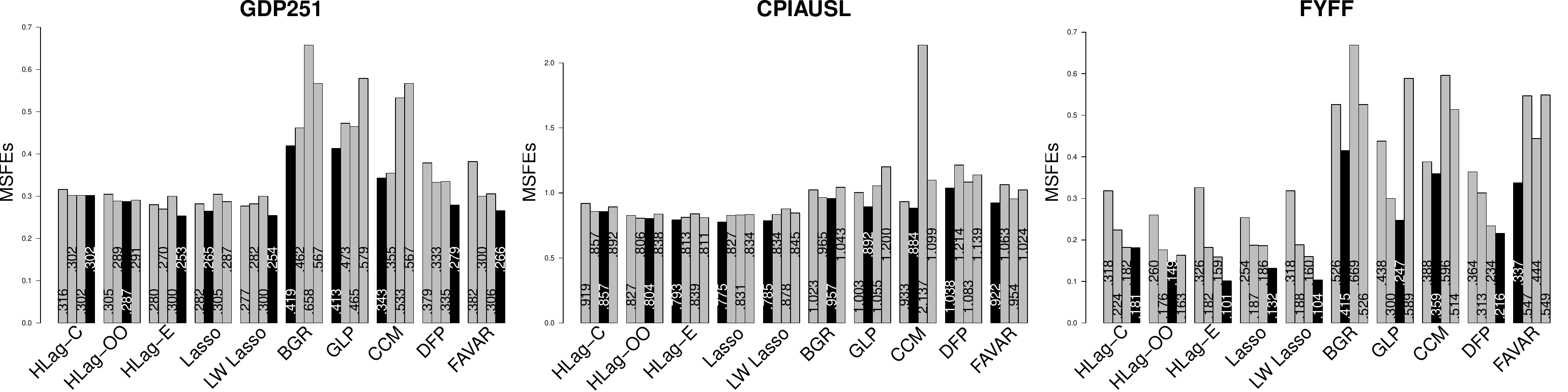}         
	\caption{ Rolling out-of-sample one-step ahead  mean squared forecast error of \textit{GDP251}, \textit{CPIAUSL} and \textit{FYFF}  for the different VAR sizes (bars from left to right: Small-Medium, Medium, Medium-Large, Large). For each method, the lowest MSFE is indicated in black. \label{MSFEmain3}}
\end{figure}

{\it Comparing forecast performance across different VAR sizes.}
To investigate whether large VARs improve forecast accuracy over smaller VARs, we turn to the MSFEs of the individual component series obtained with the multivariate forecast methods. We focus on Real Gross Domestic Product (\textit{GDP251}), 
Consumer Price Index (\textit{CPIAUSL}), and
the Federal Funds Rate (\textit{FYFF}) 
which are generally of primary interest to forecasters and policymakers. 
Figure \ref{MSFEmain3} gives the MSFEs of these three component series in the four VAR models. 

Despite the fact that the \textit{Small-Medium} VAR forecasts well for some component series, like \textit{CPIAUSL},
we often find, similar to \cite{koop}, that moving away from small VARs leads to improved forecast performance.
Consider, for instance, \textit{GDP251} and \textit{FYFF}  where half of the forecast methods give the best MSFE in the \textit{Large} VAR. 
Across the $k=10$ component series included in all four VARs, HLag, the lasso and factor methods produce the best MSFEs mainly for the \textit{Medium-Large} or \textit{Large} VARs; the  Bayesian methods mainly for the \textit{Small-Medium} or \textit{Medium} VARs.

Furthermore, the loss in forecast accuracy when adding variables to the VAR, if it occurs, remains relatively limited for 
HLag methods (on average, only 5\%) but is severe for Bayesian methods (on average,  46\%).
Although Bayesian methods perform shrinkage, all component series remain included in the larger VARs, which can severely impede forecast accuracy. HLag methods, in contrast, do not use all component series but offer the possibility to exclude possibly irrelevant or redundant variables from the forecast model.

While factor-based models produce good forecasts for larger VARs, as the factors can be estimated more precisely as the number of component series increases, 
the factors themselves do not carry, in many cases, economic interpretation. The HLag methods, in contrast, facilitate  interpretation by providing  direct insight into the component series that contribute to the good forecast performance, as discussed next.

\subsubsection{Lag Order Selection \label{interpretation}}
\begin{figure}
	\centering
	\includegraphics[width=\textwidth]{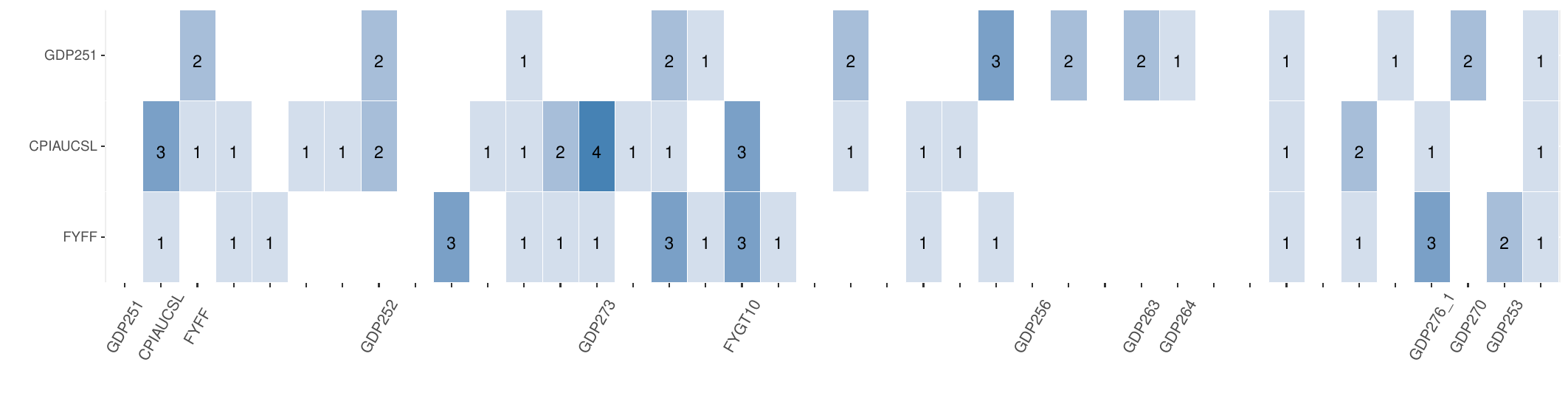} 
	\caption{\label{fig:fig7}  The first three rows of $\hat{L}^E$, the estimated elementwise maxlag matrix in the \emph{Medium-Large} VAR for the $HLag^E$ method. Components with zero maxlag are left empty.} 
\end{figure}

The HLag methods provide direct insight into the series contributing to the forecasting of each individual component. As an example, consider the 
estimated lag orders of the three main component series (\textit{GDP251}, \textit{CPIAUSL} and \textit{FYFF}) from a fitted $\text{HLag}^E_{40}$ model of the \emph{Medium-Large} group in 
Figure \ref{fig:fig7}. 
Elementwise HLag finds, for instance, that the Federal Funds Rate \textit{FYFF} is an important predictor of Gross Domestic Product  since two of its lagged components are included in the equation for forecasting \textit{GDP251}. 

Generally speaking, the lag selection results are considerably stable across time.
Figure \ref{plotsHLagE}  in Section \ref{SWAppendix} of the appendix  gives, for each end point of the rolling window, the fraction of non-zero coefficients in each of the 13 macroeconomic categories when forecasting \textit{GDP251}, \textit{CPIAUSL}, and \textit{FYFF}.
To forecast GDP growth, for instance, GDP components, employment, interest rates and stock prices have a stable and important contribution throughout the entire forecast evaluation period.

\subsection{Applications with High Sampling Rates}\label{highsampling}
The HLag methods can also be used for applications with high sampling rates. To illustrate this, we consider a financial and energy data set.

\subsubsection{Financial Application}
We apply the HLag methods to a financial data set containing realized variances for $k=16$ stock market indices, listed in Table \ref{finance_descr} of the appendix. 
Daily  realized variances based on five minute returns 
are taken from Oxford-Man Institute of 
Quantitative Finance (publicly available on  http://realized.oxford-man.ox.ac.uk/data/download).
Our data set consists of $T=4,\!163$ trading days  between January 4, 2000 and December 30, 2019. 

We compare the HLag methods to their competitors on estimated VARs with $pmax=22$ (one trading month).
The number of parameters is thus $22\times16^2+16 = 5,648$. 
December 7, 2018 to June 26, 2019 (104 observations) are used for penalty parameter selection;
June 27, 2019 to December 30, 2019 (104 observations) for forecast comparisons.

\begin{figure}     
	\centering
	\includegraphics[width=\textwidth]{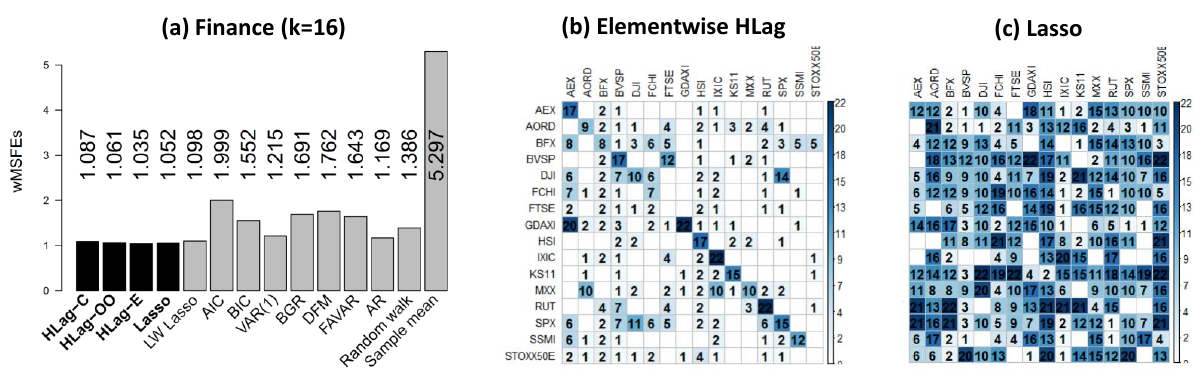}
	\caption{\label{finance_energy} Financial application. Panel (a): Rolling out-of-sample one-step-ahead $wMSFE$ with the forecast methods in the 75\% MCS  in black.
		Panel (b): Estimated  maxlag matrix for elementwise HLag and Panel (c): for the lasso.}
\end{figure}

Figure \ref{finance_energy}, panel (a) presents the one-step-ahead weighted mean squared forecast errors.\footnote{We excluded the BVAR methods GLP and CCM as they are too time consuming for large-scale VARs.}
All three HLag methods are, together with the lasso, among the best performing methods, as confirmed through the MCS. 
The HLag methods and lasso attain considerable forecast gains over all other methods.
The HLag methods' performance remains stable across different values of the maximal lag order, unlike the performance of the lasso. 
Furthermore, elementwise HLag achieves its good forecast accuracy using a more parsimonious, more interpretable description of the data than the lasso as can be seen from the estimated maxlag matrices  in Figure \ref{finance_energy} panel (b) and (c) respectively.

\subsubsection{Energy Application}
We apply the HLag methods to an energy data set \citep{candanedo2017data} containing information on $k=26$ variables related to in-house energy usage, temperature and humidity conditions. The energy data was logged every 10 minutes for about 4.5 months, giving $T=19,\!735$ observations in total.
A list of all variables and a short description is provided in Table \ref{energy_descr} of the appendix.
Data are taken from the publicly available UCI Machine Learning Repository (https://archive.ics.uci.edu/ml/data sets/Appliances+energy+prediction).

To evaluate the forecast performance of  HLag, we estimate VAR models with $pmax=6$ (one hour), thus containing $6\times26^2+26 = 4,\!082$ parameters.
May 16, 18:10 to May 17, 2016 18:00 (144 observations) are used for penalty parameter selection;
May 17, 18:10 to May 27, 2016 18:00 (1440 observations) for  forecast comparisons.

\begin{figure}[t]     
	\centering
	\includegraphics[width=\textwidth]{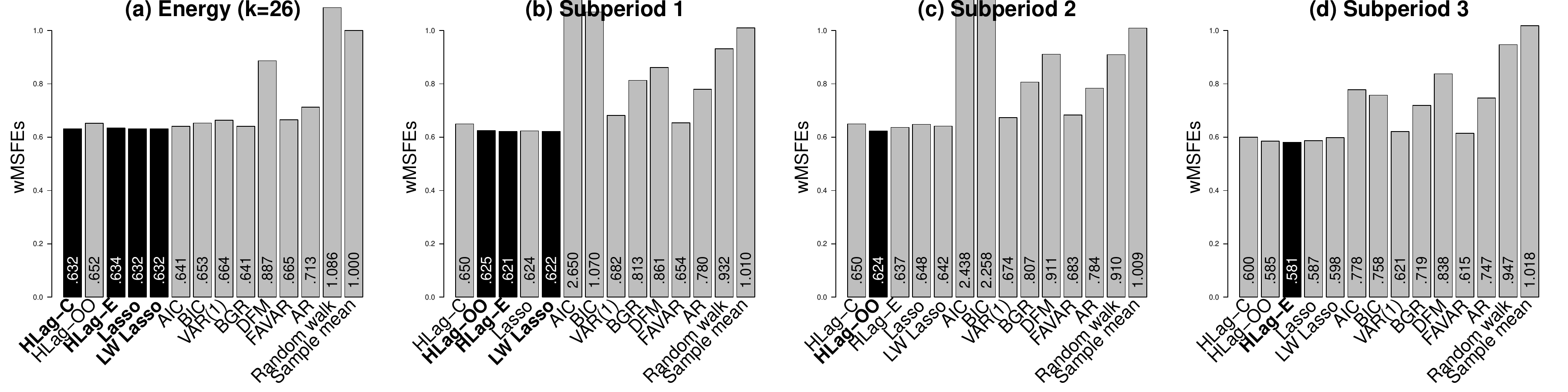} 
	
	\includegraphics[width=\textwidth]{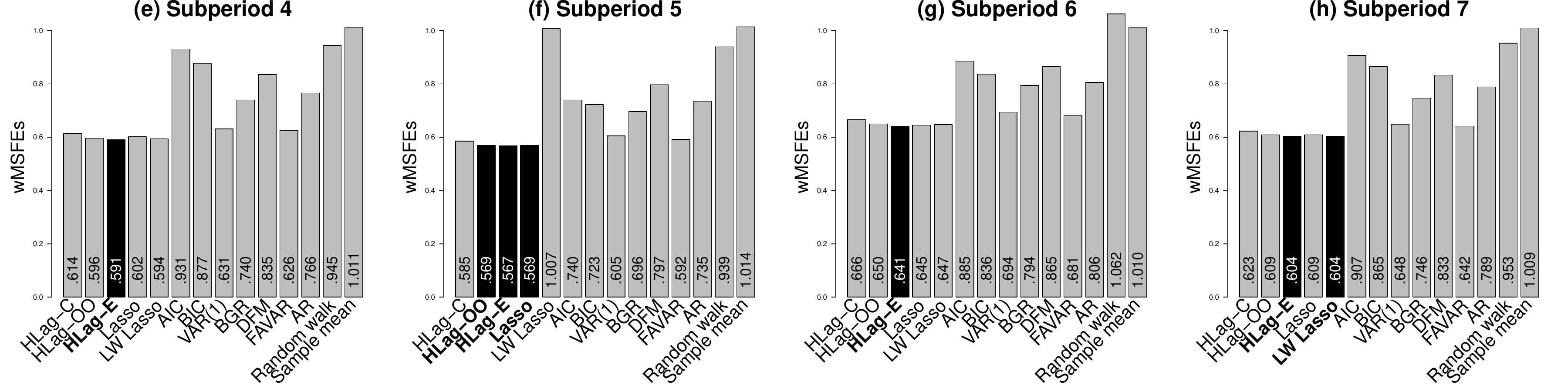} 
	
	\includegraphics[width=\textwidth]{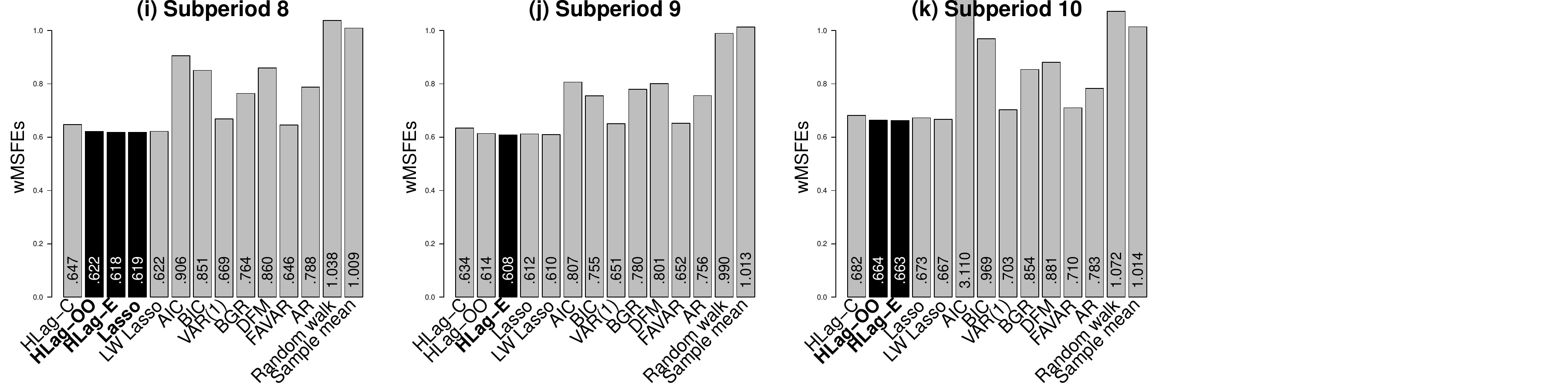} 
	\caption{\label{energy_plots} Energy application. Rolling out-of-sample one-step-ahead $wMSFE$ on the data set and on ten subperiods. Forecast methods in the 75\% MCS are indicated in black in each panel.}
\end{figure}

Figure \ref{energy_plots}  presents the one-step-ahead weighted mean squared forecast errors.\footnote{We excluded the BVAR methods GLP and CCM as they are too time consuming for large-scale VARs.}
As the sample size is large, the least squares VAR-based methods do not suffer as much from the curse of dimensionality.
Still,  HLag  has an advantage by not imposing a universal maximal lag order. On the whole data set (panel a), componentwise and elementwise HLag outperform all other methods apart from the lasso and lag-weighted lasso.
Yet, a subsample analysis reveals the dominance of elementwise HLag. 
We split the data set into ten consecutive subperiods of equal length and repeated the same forecast exercise. Results are displayed in panels (b)-(k). Elementwise HLag maintains its good performance across the subperiods and performs best. It is included in the MCS for all subperiods except for the second, 
making it a valuable addition to a forecaster's toolbox.

%% DISCUSSION %%
\section{Discussion \label{discussion}}
By incorporating the property that more recent lags convey more information than distant lags, the HLag framework offers substantial forecast improvements as well as greater insight into lag order selection than existing methods.  In addition, throughout our simulation scenarios, we see that each method is fairly robust to deviations from its particular hierarchical structure.  The substantial improvements in forecasting accuracy in data applications provide justification for the widely held belief that as the number of component series included in a model increases, the maximal lag order is not symmetric across series.  

To enforce the hierarchical lag structures, we use the nested group structure of \cite{Zhao09}. Alternatively, one could leverage the latent overlapping group lasso (LOG)  proposed by \cite{Jacob09}.
While \cite{Yan2017} indicate that the nested group structures might suffer from a more aggressive shrinkage of parameters deep in the hierarchy (i.e. higher-order autoregressive coefficients), in the VAR model, large amounts of shrinkage on the more distant lags versus small amounts of shrinkage on the more recent lags may be desirable \citep{BickelSong}. In our simulation studies, the nested group lasso structures significantly outperformed the LOG structures in the large majority of cases. Especially as the maximal lag order increases, the nested group lasso turned out to be more robust. Detailed results are available in Section \ref{LOG} of the appendix. 

Implementations of our methods are available in the R package {\tt BigVAR}, which is hosted on the Comprehensive R Archive Network (cran). 
Despite the more challenging computational nature of overlapping group lasso problems compared to conventional sparsity or non-overlapping group sparsity problems (e.g., \citealp{chen20141}, \citealp{yuan2011efficient}, \citealp{mairal2010network}),
our methods scale well and are computationally feasible in high dimensions.  
For instance, for the \textit{Large} VAR ($k=168, T=195,$ and $113,\!064$ parameters)  estimated on the Stock and Watson data,
the HLag methods only require (on an Intel Xean Gold 6126 CPU @ 2.60GHz machine) around 
1.5  (Own-Other), 2  (Componentwise) and 3.5 minutes (Elementwise), 
including penalty parameter selection.  This requires estimating the VAR  610 times ($61 \text{ time points}\times10\text{ penalty parameters}$).  For fixed penalty parameter, the HLag methods can be computed in less than a second. 
The computational bottleneck of our implementation thus concerns the penalty parameter selection. Alternatives (information criteria or a time series cross-validation search where the models are not re-estimated  every single time point but at a lower sampling frequency) can be considered to reduce the penalty parameter search for applications with high sampling rates.
To be widely adopted by 
practitioners, we do think that our methods have a considerable advantage compared to more computationally intensive methods such as the lag-weighted lasso, the Bayesian CCM and GLP approaches requiring around  
33 minutes (Lag-weighted lasso) or even more than 2 hours (Bayesian methods) for one model fit of the \textit{Large} Stock and Watson VAR. 
At the very least, one of the proposed HLag approaches can be quickly run to provide numerous insights before a more computationally demanding method is adopted. 

The HLag framework is quite flexible and can  be extended in various ways. 
For example, more complicated weighting schemes (see e.g., \citealp{jenatton2011structured}, \citealp{bien2016convex})  could be adopted to address the more aggressive shrinkage of parameters deep in the hierarchy, but these make computation more involved \citep{Yan2017} and our simulations in Section \ref{LOG} of the appendix indicate that this may not be beneficial in the VAR setting.
Furthermore, if the practitioner prefers to summarize the information content in large data sets by constructing few factors, HLag penalties can, for instance, be applied with minimal adaption to the factors augmenting the VAR in a FAVAR.
The HLag framework would allow one to flexibly vary the number of factors in each marginal equation of the FAVAR and to automatically determine the lag order of the factors, in addition to the lag structure of the autoregressive components.
Finally, building on \cite{basu2013estimation}, we derive preliminary theoretical results on prediction consistency for HLag in a high-dimensional regime. Given the complicated nested group structure of the HLag penalty,  work is needed to further explore its theoretical properties. To this end, recent advances in the theory of the hierarchical group lasso (e.g., \citealp{Yan2017, Yu2017}) could be leveraged.

\appendix
%% APPENDIX %%
\section{Theoretical properties: Proofs}\label{theory}
We start by proving two auxiliary results, then we combine these in the proof of Theorem \ref{mainresult}. For ease of notation and without loss of generality, we omit the intercept vector $\utwi{\nu}$ from the VAR model \eqref{VAR1a}.
\begin{lem}
	If $\lambda\ge
	\max_{\ell}\|\frac1{T}\sum_{t=1}^T\mathbf{y}_{t-\ell}\mathbf{u}_t^\top\|_{\infty}$,
	then 
	\begin{align*}
	\frac1{2T}\sum_{t=1}^{T}\|\sum_{\ell=1}^p(\PhiB^{(l)}-\hat{\PhiB}^{(l)})
	\mathbf{y}_{t-\ell}\|^2_2&\le
	2\lambda{\cal P}^E_{\text{HLag}}({\bf{\PhiB}}).
	\end{align*}
	\label{lem:slowrate}
\end{lem}

\begin{proof}{\textbf{of Lemma \ref{lem:slowrate}.}}
	Since $\hat{\PhiB}$ is a minimizer of \eqref{optim}, we have that
	$$
	\frac1{2T}\sum_{t=1}^{T}\|\mathbf{y}_t-\sum_{\ell=1}^p\hat{\PhiB}^{(l)}
	\mathbf{y}_{t-\ell}\|^2_2+\lambda{\cal P}^E_{\text{HLag}}({\bf \hat{\PhiB}})
	\le 
	\frac1{2T}\sum_{t=1}^{T}\|\mathbf{y}_t-\sum_{\ell=1}^p \PhiB^{(\ell)}\mathbf{y}_{t-\ell}\|^2_2+\lambda{\cal P}^E_{\text{HLag}}(\PhiB).
	$$
	Substituting the data generating process $\mathbf{y}_t = \PhiB^{(1)}\mathbf{y}_{t-1}+\dots+\PhiB^{(p)}\mathbf{y}_{t-p}+\mathbf{u}_t$ into the above, we obtain 
	$$
	\frac1{2T}\sum_{t=1}^{T}\|\mathbf{u}_t+\sum_{\ell=1}^p(\PhiB^{(l)}-\hat{\PhiB}^{(l)})
	\mathbf{y}_{t-\ell}\|^2_2+\lambda{\cal P}^E_{\text{HLag}}({\bf \hat{\PhiB}})
	\le 
	\frac1{2T}\sum_{t=1}^{T}\|\mathbf{u}_t 
	\|^2_2+\lambda{\cal P}^E_{\text{HLag}}(\PhiB).
	$$
	After re-arranging, we get
	\begin{align*}
	\frac1{2T}\sum_{t=1}^{T}\|\sum_{\ell=1}^p(\PhiB^{(l)}-\hat{\PhiB}^{(l)})
	\mathbf{y}_{t-\ell}\|^2_2+\lambda{\cal P}^E_{\text{HLag}}({\bf \hat{\PhiB}})&\le
	\frac1{T}\sum_{t=1}^T
	\sum_{\ell=1}^p\mathbf{u}_t^\top(\hat{\PhiB}^{(l)}-\PhiB^{(l)}) y_{t-\ell}+\lambda{\cal P}^E_{\text{HLag}}({\bf{\PhiB}}).
	\end{align*}
	Now,
	\begin{align*}
	\frac1{T}\sum_{t=1}^T
	\sum_{\ell=1}^p\mathbf{u}_t^\top(\hat{\PhiB}^{(l)}-\PhiB^{(l)}) \mathbf{y}_{t-\ell} 
	=&\frac1{T}\sum_{\ell=1}^p \langle \hat{\PhiB}^{(l)}-\PhiB^{(l)}, \sum_{t=1}^T \mathbf{y}_{t-\ell}\mathbf{u}_t^\top\rangle\\
	& \qquad \le \frac1{T}\|\hat{\PhiB}-\PhiB\|_1\max_{\ell}\|\sum_{t=1}^T\mathbf{y}_{t-\ell}\mathbf{u}_t^\top\|_{\infty} \\
	&\qquad \le \frac1{T}{\cal P}^E_{\text{HLag}}(\hat{\PhiB} - \PhiB)
	\max_{\ell}\|\sum_{t=1}^T\mathbf{y}_{t-\ell}\mathbf{u}_t^\top\|_{\infty}
	\end{align*}
	since $\|\PhiB\|_1 := \sum_{i=1}^k\sum_{j=1}^k\sum_{\ell=1}^p\|\PhiB_{ij}\|_1 \leq \sum_{i=1}^k\sum_{j=1}^k\sum_{\ell=1}^p\|\PhiB_{ij}^{(\ell:p)}\|_2 := {\cal P}^E_{\text{HLag}}(\PhiB).$
	Thus,
	\begin{align*}
	\scriptsize
	\frac1{2T}\sum_{t=1}^{T}\|\sum_{\ell=1}^p(\PhiB^{(l)}-\hat{\PhiB}^{(l)})
	\mathbf{y}_{t-\ell}\|^2_2+\lambda{\cal P}^E_{\text{HLag}}({\bf \hat{\PhiB}})&\le
	\frac1{T}{\cal P}^E_{\text{HLag}}(\hat{\PhiB} - \PhiB)
	\max_{\ell}\|\sum_{t=1}^T\mathbf{y}_{t-\ell}\mathbf{u}_t^\top\|_{\infty} + \lambda{\cal P}^E_{\text{HLag}}({\bf{\PhiB}}).
	\end{align*}  
	Under the assumption on $\lambda$, we get
	\begin{align*}
	\frac1{2T}\sum_{t=1}^{T}\|\sum_{\ell=1}^p(\PhiB^{(l)}-\hat{\PhiB}^{(l)})
	\mathbf{y}_{t-\ell}\|^2_2+\lambda{\cal P}^E_{\text{HLag}}({\bf \hat{\PhiB}})&\le
	\lambda {\cal P}^E_{\text{HLag}}(\hat{\PhiB} - \PhiB)
	+ \lambda{\cal P}^E_{\text{HLag}}({\bf{\PhiB}}).
	\end{align*} 
	The result follows from observing that ${\cal P}^E_{\text{HLag}}(\hat{\PhiB} - \PhiB) \leq {\cal P}^E_{\text{HLag}}(\hat{\PhiB}) + {\cal P}^E_{\text{HLag}}(\PhiB)$.
\end{proof}

\begin{lem} \label{lemma2}
	If $T>25\log(pk^2)$, $T>4$,
	$pk^2\gg1$ 
	and we choose $\lambda\ge30v(\PhiB,\mathbf{\Sigma}_u)\sqrt{\log(pk^2)/T}$, then
	$$
	\max_{\ell,j,k}|\frac1{T}\sum_{t=1}^Ty_{t-\ell,j}u_{t,k}|\le\lambda
	$$
	with probability at least $1-\frac{12}{(pk^2)^{23/2}}$.
\end{lem}

\begin{proof}{\textbf{of Lemma \ref{lemma2}.}}
	In the middle of page 20 of Basu and Michailidis (2013)\footnote{S. Basu and G. Michailidis. Estimation in High-dimensional Vector Autoregressive
		Models. arXiv:1311.4175v1, 2013.} (a preliminary version of \citealp{basu2013estimation}),
	it is shown that, 
	$$
	\P\left(\frac{1}{T}\max_{\ell,j,k}|\sum_{t=1}^Ty_{t-\ell,j}u_{t,k}|>b\right)\le12\exp\left\{-\frac{T}{2}\min\left\{1,\left(\frac{b}{6v(\PhiB,\mathbf{\Sigma}_u)}-\frac{2}{\sqrt{T}}\right)^2\right\}+\log(pk^2)\right\}
	$$
	where 
	$$
	b=(18+6\sqrt{2(A+1)}) v(\PhiB,\mathbf{\Sigma}_u)\sqrt{\log(pk^2)/T}
	$$
	for some constant $A>0$. For simplicity, we take $A=1$. 
	Note that the exponent can be written as 
	\begin{eqnarray}
	{\footnotesize
		-\frac{T}{2}\min\left\{1,\left(\frac{b}{6v(\PhiB,\mathbf{\Sigma}_u)}-\frac{2}{\sqrt{T}}\right)^2\right\}+\log(pk^2)}&=&{\footnotesize -\frac{T}{2}\min\left\{1,\left(\frac{5\sqrt{\log(pk^2)}-2}{\sqrt{T}}\right)^2\right\}+\log(pk^2)} \nonumber \\
	& = & {\footnotesize -\frac{1}{2}\min\left\{T,\left(5\sqrt{\log(pk^2)}-2\right)^2\right\}+\log(pk^2).} \nonumber
	\end{eqnarray}
	Since $T>25\log(pk^2)$ and $T>4$, it follows that
	$
	T>\left(5\sqrt{\log(pk^2)}-2\right)^2
	$
	and the exponent is
	\begin{align*}
	-\frac{1}{2}[5\sqrt{\log(pk^2)}-2]^2+\log(pk^2)=-(23/2)\log(pk^2)+10\sqrt{\log(pk^2)}-2\approx -(23/2)\log(pk^2),
	\end{align*}
	where the last approximation follows from the assumption that $pk^2\gg1$.
	
	Thus, for the choice of $\lambda$ given above, we have that 
	$$
	\P\left(\max_{\ell,j,k}|\sum_{t=1}^Ty_{t-\ell,j}u_{t,k}|\le\lambda\right)\ge1-12\exp\left\{-(23/2)\log(pk^2)\right\}=1-\frac{12}{(pk^2)^{23/2}}.
	$$
\end{proof}

\begin{proof}{\textbf{of Theorem \ref{mainresult}.}}
	Combining  the results from Lemma \ref{lem:slowrate} and \ref{lemma2}, we have for  $T>25\log(pk^2)$, $T>4$, $pk^2\gg1$ and
	$\lambda\ge30v(\PhiB,\mathbf{\Sigma}_u)\sqrt{\log(pk^2)/T}$, that
	$$
	\msfein\le\tr(\mathbf{\Sigma}_u)+ 
	4\lambda{\cal P}^E_{\text{HLag}}({\bf{\PhiB}})
	$$ or alternatively
	$$
	\frac1{T}\sum_{t=1}^T \left\|\sum_{\ell=1}^p(\hat\PhiB^{(\ell)} - \PhiB^{(\ell)})\mathbf{y}_{t-\ell}\right\|^2_2 \leq 
	4\lambda{\cal P}^E_{\text{HLag}}({\bf{\PhiB}})
	$$
	with probability at least $1-\frac{12}{(pk^2)^{23/2}}$.
	
	Assuming that all coefficients are bounded by $M$, we have that 
	$$\|\PhiB_{ij}^{(\ell:L_{ij})}\|_2\le M\sqrt{L_{ij}-\ell+1}\le M\sqrt{L_{ij}},$$
	so
	$$
	{\cal P}^E_{\text{HLag}}({\bf{\PhiB}}) = \sum_{i=1}^k\sum_{j=1}^k\sum_{\ell=1}^{L_{ij}}\|\PhiB_{ij}^{(\ell:L_{ij})}\|_2\le  M\sum_{i=1}^k\sum_{j=1}^kL_{ij}^{3/2},
	$$
	and finally, 
	$$
	\frac1{T}\sum_{t=1}^T \left\|\sum_{\ell=1}^p(\hat\PhiB^{(\ell)} - \PhiB^{(\ell)})\mathbf{y}_{t-\ell}\right\|^2_2 
	\lesssim M v(\PhiB, \mathbf{\Sigma}_u)\sqrt{\frac{\log(pk^2)}{T}}\sum_{i=1}^k\sum_{j=1}^kL_{ij}^{3/2}.
	$$
\end{proof}

\section{Comparison Methods \label{benchmarks}}
\subsection{Least Squares VAR} A standard method in lower dimensional settings is to fit a $VAR_k(\ell)$
with least squares for $0\le\ell\le pmax$ and then to select a universal lag order $\ell$ using AIC or BIC.
Per \cite{Lutk}, the AIC and BIC of a $VAR_k(\ell)$ are defined as
\begin{align}
\nonumber
&\text{AIC}(\ell)=\log\det(\hat{\bf\Sigma}_u^{\ell})+ \frac{2k^2\ell}{T},\\
&\text{BIC}(\ell)=\log\det(\hat{\bf\Sigma}_u^{\ell})+ \frac{\log(T)k^2\ell}{T},  \nonumber 
\end{align}
in which $\hat{\bf\Sigma}_u^{\ell}$ is the residual sample covariance matrix having
used least squares to fit the $VAR_k(\ell)$.  The
lag order $\ell$ that minimizes $\text{AIC}(\ell)$ or $\text{BIC}(\ell)$ is selected.  This
method of lag order selection is only possible when $k\ell\le T$ since
otherwise least squares is not well-defined.  In simulation Scenarios 1-3 ($T=100$), we cannot use least squares for $\ell>1$, thus for a simple
benchmark we instead estimate a $VAR_k(1)$ by least squares:
$$
\min_{{\boldsymbol\nu},\PhiB}\left\{\frac{1}{2T}\|\Y-{\boldsymbol\nu}{\bf 1}^\top-\PhiB^{(1)}\Z^{(1)}\|_2^2\right\},
$$
where $\Z^{(1)}=[\bf y_0~\cdots~\bf y_{T-1}]$.

\subsection{Lasso VAR} We also include two well-known lasso-based VAR regularization approaches. The
{\em lasso} estimates the VAR using an $L_1$-penalty:
$$
\min_{{\boldsymbol\nu},\PhiB}\left\{\frac{1}{2T}\|\Y-{\boldsymbol\nu}{\bf 1}^\top-\PhiB\Z\|_2^2 +\lambda\|\PhiB\|_1\right\},
$$
where $\|\PhiB\|_1$ denotes $\|\text{vec}(\PhiB)\|_1$.
The lasso does not intrinsically consider lag order, hence
\citet{BickelSong} propose a {\em lag-weighted lasso}
penalty in which a weighted $L_1$-penalty is used with weights that increase geometrically with lag order:
$$
\min_{{\boldsymbol\nu},\PhiB}\left\{\frac{1}{2T}\|\Y-{\boldsymbol\nu}{\bf 1}^\top-\PhiB\Z\|_2^2+\lambda\sum_{\ell=1}^p\ell^\alpha\|\PhiB^{(\ell)}\|_1\right\}.
$$
The tuning parameter $\alpha\in [0,1]$ determines how fast the
penalty weight increases with lag.  While this form of penalty applies
greater regularization to higher order lags, it is less structured than our HLag
penalties in that it does not necessarily produce sparsity patterns in which all coefficients
beyond a certain lag order are zero. The regularization parameters $\lambda$ and $\alpha$ are jointly selected using a two-dimensional penalty parameter search. We have implemented these methods in \texttt{R}, the code is available as Supplementary Material. 

\subsection{Bayesian VAR} We consider three Bayesian benchmarks: the method of \cite{BGR}, \cite{GLP} and \cite{Carriero17}. 
These approaches are also applicable to a situation like ours where many parameters need to be estimated but the observation period is limited.
However, in contrast to the HLag methods, these methods are not sparse (parameter estimates are only shrunken towards zero) and do not perform lag order selection. 

\cite{BGR} use a modified Minnesota prior which leads to a posterior for the autoregressive parameters, conditional on the error variance-covariance matrix, that is normal.  
As we transformed all variables for
stationarity, we set all prior means in the BGR implementation to zeros. Following \cite{BGR}, we select the hyperparameter that controls the degree of regularization as that which minimizes the $h$-step ahead MSFE across the $k$ component series. We have implemented this method in \texttt{R}, the code is available as Supplementary Material.

\cite{GLP} choose the informativeness of the priors in an ``optimal" way by treating the priors as additional parameters, as in hierarchical modeling. 
We use the authors' replication files (\texttt{Matlab}-code) publicly available at https://www.newyorkfed.org/research/economists/giannone/pub. 

\cite{Carriero17} use a  general Minnesota-based independent prior to allow for a more flexible lag choice. 
Note that the authors also allow for stochastic volatility, but we compare the HLag methods to their ``homoscedastic" BVAR  that does not allow for stochastic volatility, in line with the other methods considered in this paper. We adapt the authors' code (publicly available at  http://didattica.unibocconi.eu/mypage/index.php?IdUte=49257\&idr=27515\&lingua=eng) to this homoscedastic setting by combining it with \texttt{Matlab} code for BVAR using Gibbs sampling available at https://sites.google.com/site/dimitriskorobilis/matlab/code-for-vars.
For full technical details on the Bayesian methods, we refer the reader to \cite{BGR}, \cite{GLP} and \cite{Carriero17}  respectively.

\subsection{Factor Models}
We consider two factor-based benchmarks: a Dynamic Factor Model (DFM, see e.g. \citealp{forni2000generalized}; \cite{stock2002forecasting}) and a Factor Augmented VAR Model (FAVAR, \citealp{Bernanke05}). In contrast to the HLag methods, these methods do not achieve dimension reduction by sparsity. Instead, the information contained in the large predictor set is summarized by few factors. We estimate the factors by Principal Component Analysis and follow \cite{McCracken2016} in using the $PC_{p2}$ criterion, developed in \cite{Bai2002}, to select the number of factors

Regarding the DFM, the time series are regressed on lagged values of the factors. The factors are obtained from the whole data set and their lag order is determined via AIC. Similar results are obtained with BIC and  available from the authors upon request. 
Regarding the FAVAR model, we regress each time series on its own lagged values and lagged values of the factors. The factors are obtained from the data set of all other variables. Lag selection is done via AIC, while similar results are obtained with BIC.
We have implemented both methods in \texttt{R}, the code is available as Supplementary Material.

\subsection{Other Methods}
Finally, we compare against three simple
baselines. The unconditional {\em sample mean} corresponds to the
intercept-only model,
$$
\min_{{\boldsymbol\nu}}\frac{1}{2T}\|\Y-{\boldsymbol\nu}{\bf 1}^\top\|_2^2,
$$
which makes one-step-ahead forecasts of the form
$\hat{\bf y}_{t+1}=\frac1{t}\sum_{\ell=1}^{t}{\bf y}_{\ell}$. The vector {\em random walk} model, which corresponds to 
$$\hat{\boldsymbol\nu}={\bf 0},\qquad \hat{\PhiB}^{(1)}={\bf I}_k
,\qquad \hat{\PhiB}^{(2:p)}={\bf 0},$$
and makes one-step-ahead forecasts of the form
$\hat{{\bf y}}_{t+1}={\bf y}_{t}$.
Finally, we consider a separate autoregressive model for each time series. To simultaneously obtain parameter estimates and select the lag order, we use the univariate analogue of equation \eqref{optim}
\begin{align}
\min_{\phi_i}\left\{\frac1{2T}\|{\bf Y}_i-\boldsymbol\phi_i{\bf X}\|_2^2+\lambda_i  \sum_{\ell=1}^{p} ||\boldsymbol\phi_i^{(\ell:p)}||_2\right\}. \nonumber
\end{align}
for each component series $i=1,\ldots,k$ with ${\bf X} = [\mathbf{x}_1~\cdots~\mathbf{x}_{T}] \in\mathbb R^{p\times T}$, $\mathbf{x}_t = [y_{i, t-1}~\cdots~y_{i,t-p}]^\top$ $\in\mathbb R^{p\times 1}$ and $\phi_i\in\mathbb R^{1\times p}$. As such, the univariate AR  is a special univariate case of the multivariate elementwise HLag  introduced in Section \ref{sec:lag-hier-group}. For each individual autoregression, we take the maximal autoregressive order equal to the true VAR order $p$ in the simulations. In the empirical application we take four as maximal autoregressive order.

\section{Simulation Study}
\subsection{Simulation Scenarios }
\label{SimScenarios}

{\it Simulation Scenario 1: Componentwise Lag Structure.}
In this scenario, we simulate according to an HLag$^C_{45}(5)$
structure.  In particular, we choose the maxlag matrix
$$\LL=
[1,2,3,4,5]^\top\otimes({\bf 1}_{9}{\bf 1}_{45}^\top).
$$
This $45\times 45$ maxlag matrix is row-wise constant, meaning that all components
{\em within a row} have the same maxlag; we partition the rows into 5
groups of size 9, each group taking on a distinct maxlag in
$\{1,2,3,4,5\}$.  A coefficient matrix $\PhiB$ with maxlag matrix
$\LL$ is used in Scenario 1's simulations and its magnitudes are depicted in Figure 4, panel (1) of the manuscript.

{\it Simulation Scenario 2: Own-Other Lag Structure.}
In this scenario, we create the matrix $\PhiB$ in such a manner that
it differentiates between \emph{own} and \emph{other} coefficients.  The
coefficients of a series' ``own lags'' (i.e., $\PhiB_{ii}^{(\ell)}$) are
larger in magnitude than those of ``other lags''
(i.e., $\PhiB_{ij}^{(\ell)}$ with $i\neq j$).  The magnitude of
coefficients decreases as the lag order increases.  The
HLag$^O_{45}(2)$ model we simulate is depicted in Figure
4, panel (2) of the manuscript. The first 15 rows
can be viewed as univariate autoregressive models in which only the
\emph{own} term is nonzero; in the next 15 rows, for the first $k$
coefficients, the coefficient on a series' \emph{own} lags is larger than
``other lags,'' and, for the next $k$ coefficients, only \emph{own}
coefficients are nonzero; the final 15 rows have nonzeros throughout
the first $2k$ coefficients, with \emph{own} coefficients dominating
\emph{other} coefficients in magnitude.

{\it Simulation Scenario 3: Elementwise Lag Structure.}
In this scenario, we simulate under an HLag$^E_{45}(4)$ model,
meaning that the maxlag is allowed to vary not just across rows but also
{\em within} rows.  Each marginal series in each row is randomly assigned a maxlag of either 1 (with 90 percent probability) or 4 (with 10 percent probability).  The coefficient matrices are depicted in Figure 4, panel (3).

{\it Simulation Scenario 4: Data-based Lag Structure.}
Similar to \cite{carriero2012}, we carry out a simulation by bootstrapping the actual \textit{Medium-Large} macroeconomic data set with $k=40$ and $T=195$ as discussed in Section \ref{Sec5} of the manuscript.
We start from the estimates obtained by applying the Bayesian approach of \cite{GLP} to this data set with $pmax=4$. The obtained estimates of the autoregressive matrices are visualized in Figure 4, panel (4) and the autoregressive matrices verify the VAR stability conditions. 
We then construct our simulated data using a non-parametric residual bootstrap procedure (e.g., \citealp{Kreiss12}) with bootstrap errors an i.i.d.\ sequence of discrete random variables uniformly distributed on $\{1, \ldots, T\}$.

\subsection{Generation of Simulation Scenarios}
\label{SimScenGen}
All of our simulation structures were generated to ensure a stationary coefficient matrix, $\PhiB$.  In order to construct a coefficient matrix for these scenarios, we started by converting the $\text{VAR}_{k}(p)$ to a $\text{VAR}_{k}(1)$ as described in equation 2.1.8 of \cite{Lutk}
\begin{align}
\label{bigcm}
\mathbf{A}=
\begin{bmatrix}
\PhiB^{(1)} & \PhiB^{(2)} & \dots & \PhiB^{(p-1)} & \PhiB^{(p)}\\
\utwi{I}_{k}& \utwi{0} &\utwi{0} &\utwi{0}&\utwi{0}\\
\utwi{0}& \utwi{I}_{k} &\utwi{0} &\utwi{0}&\utwi{0}\\
\vdots & \vdots & \ddots & \vdots &\vdots\\ 
\utwi{0}& \utwi{0} &\utwi{0} &\utwi{I}_k&\utwi{0}
\end{bmatrix}
\end{align}
For $\mathbf{A}$ to be stationary, its maximum eigenvalue must be less than 1 in modulus.  In general, it is very difficult to generate stationary coefficient matrices.  \cite{Bosh} offer a potentially viable procedure that utilizes the unique structure of equation \eqref{bigcm}, but it does not allow for structured sparsity.  We instead follow the approach put forth by \cite{gilbert} in which structured random coefficient matrices are generated until a stationary matrix is recovered. 

\subsection{Sensitivity Analysis: Choice of Group Lasso Formulation \label{LOG}}
The hierarchical lag structures of the HLag methods can either be enforced via the nested group structure of \cite{Zhao09} or via the latent overlapping group lasso (LOG) proposed by \cite{Jacob09}.
We compare the LOG to the nested group structures in our simulation studies. 

\begin{table}
	\resizebox{0.8\textwidth}{!}{\begin{minipage}{\textwidth}
			\begin{tabular}{llllccccccc} \hline 
				&&&& \multicolumn{5}{c}{Simulation Scenarios}	\\
				Order&HLag 		    &&& 1. Componentwise & 2. Own-Other & 3. Elementwise & 4. Data & 5. Robustness  \\ \hline 
				Known&Componentwise   &&& \textbf{1.036} & 0.980 & 0.946 & \textbf{1.019} & 1.004\\
				&Own-other       &&& \textbf{1.048} & 1.000 & 0.944 & 1.006 & 0.986\\
				&Elementwise     &&& \textbf{1.037} & 1.001 & 0.944 & \textbf{1.010} & 1.005\\ 
				Unkown&Componentwise    &&& \textbf{1.138}& \textbf{1.235}& \textbf{1.094}& \textbf{1.051}& \textbf{1.030}\\
				&Own-other              &&& \textbf{1.119}& \textbf{1.171}& \textbf{1.064}& \textbf{1.031}& \textbf{1.014}\\
				&Elementwise            &&& \textbf{1.053}& \textbf{1.142}& \textbf{1.030}& \textbf{1.028}& \textbf{1.037}\\ \hline 
			\end{tabular}
	\end{minipage} }
	\caption{Out-of-sample mean squared forecast errors  of the LOG  relative to that of the nested group lasso in  Scenario 1 to 5. Outperformance (as confirmed with paired $t$-tests) by the nested group lasso is indicated in bold. \label{LOG_MSFEs}}
\end{table}

\begin{table}
	\resizebox{0.8\textwidth}{!}{\begin{minipage}{\textwidth}
			\begin{tabular}{llllccccccc} \hline 
				&&&& \multicolumn{5}{c}{Simulation Scenarios}	\\
				Measure&HLag 		    &&& 1. Componentwise & 2. Own-Other & 3. Elementwise & 4. Data & 5. Robustness  \\ \hline 
				$L_1$-lag&Componentwise    &&&  \textbf{4.581} & \textbf{7.072} & \textbf{3.145} & 1.002 & \textbf{1.211}\\
				error&Own-other         &&&  \textbf{4.837} & \textbf{3.271}& \textbf{3.076} &  1.001 &\textbf{1.177} \\
				&Elementwise               &&&  \textbf{1.529} & \textbf{2.342}& \textbf{1.383} &  1.001& \textbf{1.058}\\ \hline 
				
				$L_\infty$-lag&Componentwise    &&&  \textbf{2.418} & \textbf{3.150} & \textbf{1.752} & \textbf{1.281} & \textbf{1.203}\\
				error&Own-other              &&&  \textbf{2.513} & \textbf{4.979} & \textbf{1.780} & \textbf{1.138} & \textbf{1.741}\\
				&Elementwise                    &&&  \textbf{1.015} & \textbf{3.091}& \textbf{1.683}  & \textbf{1.636} & \textbf{1.266}\\ \hline 
			\end{tabular}
	\end{minipage} }
	\caption{Lag order selection   of the LOG  relative to that of the nested group lasso in  Scenario 1 to 5. Outperformance (as confirmed with paired $t$-tests) by the nested group lasso is indicated in bold.\label{LOG_lags}}
\end{table}

In Table \ref{LOG_MSFEs}, we present the MSFEs of the  LOG  structures relative to those of the nested group lasso (for each HLag method). 
In Table \ref{LOG_lags} their lag order selection performance is compared.
Values above one indicate better performance of the nested group lasso compared to the LOG. 
In both Tables, the nested group lasso significantly outperforms the LOG in the vast majority of cases. 
Especially when the maximal lag order $pmax$ increases, the nested group lasso structures perform better than the LOG structures. 

The finding that the nested group lasso structures are more robust than the LOG structures as  $pmax$ increases, is confirmed through the Robustness simulation scenario. In Table \ref{LOG_MSFEs_robustness}, we report the MSFEs and lag order  measures as $pmax$ increases from its true order (five) to $pmax=50$. On all performance measures, the nested group lasso structures perform, overall, better than the LOG structures and the margin by which the former outperforms the latter increases with $pmax$.

\begin{table}
	\centering
	\begin{tabular}{llllccccc} \hline 
		Performance &&&& \multicolumn{5}{c}{Maximal lag order}	\\
		measure&HLag 		    & &&& $pmax = 5$ & $pmax = 12$ & $pmax = 25$ & $pmax = 50$  \\ \hline 
		MSFE &Componentwise   & &&& 1.004 & \textbf{1.030} & \textbf{1.061} & \textbf{1.114} \\
		&Own-other    & &&& 0.986 & \textbf{1.014} & \textbf{1.049} & \textbf{1.105}\\ 
		&Elementwise  & &&& 1.005 & \textbf{1.037} & \textbf{1.075} & \textbf{1.151} \\ 
		&&&&  & &  &&  \\
		$L_1$-lag&Componentwise &&&&  0.912  & \textbf{1.211}  & \textbf{1.331} & \textbf{1.969}\\
		error &Own-other     &&&&  \textbf{1.090}  & \textbf{1.177}  & \textbf{1.226} & \textbf{1.529}\\
		&Elementwise           &&&&  0.990  & \textbf{1.058} & \textbf{1.100}  & \textbf{1.149}\\
		&&&&  & &  &&  \\
		$L_\infty$-lag&Componentwise    &&&& 0.985 & \textbf{1.203} & \textbf{1.399} & \textbf{2.491}\\
		error& Own-other             &&&& \textbf{1.127} & \textbf{1.741}  & \textbf{2.263} & \textbf{4.130}\\
		&Elementwise                   &&&& 1.000 & \textbf{1.266} & \textbf{1.898} &\textbf{2.630}\\ \hline 
	\end{tabular}
	\caption{Robustness simulation scenario: Forecast performance and lag order selection  of the LOG  relative to that of the nested group lasso for different values of the maximal lag order  $pmax$. Outperformance (as confirmed with paired $t$-tests) by the nested group lasso is indicated in bold.\label{LOG_MSFEs_robustness}}
\end{table}	

\subsection{Sensitivity Analysis: Impact of Increasing the Time Series Length \label{increaseT}}
\begin{figure}
	\centering	
	\includegraphics[width=6.9cm]{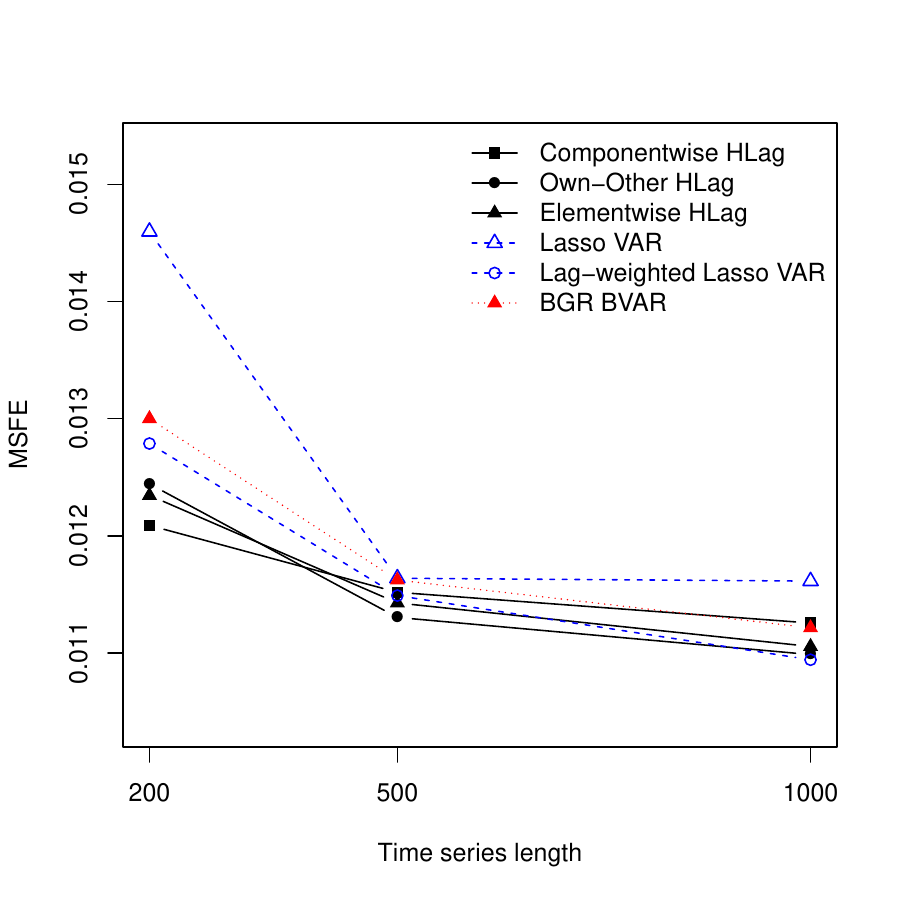} 	
	\caption{Robustness simulation scenario: Out-of-sample  mean squared  forecast errors, for different values of the sample size $T$. Note that we have not included the BVAR methods GLP and CCL as they are too time consuming for large-scale VARs.}\label{Sim4MSFE_increaseT}
\end{figure}

We investigate the impact of increasing the time series length on our forecast accuracy results. We use the autoregressive parameter structure of Scenario 5 and increase the time series length from from $T=200$ over $T=500$ to $T=1000$ while keeping the maximal lag order $pmax=5$.
Figure \ref{Sim4MSFE_increaseT}  presents the MSFEs.
The forecast errors of
all methods decrease as $T$ increases, in line with our expectations. While the difference between the methods decreases as the sample size increases, all HLag methods sill significantly outperform the lasso.

\subsection{Sensitivity Analysis: Choice of Error Covariance matrix \label{errorcov}}
We investigate the sensitivity of our forecast accuracy results to the choice of error covariance matrix. We start from the autoregressive parameter structure of Scenario 5 ($pmax=5$) and  consider, in turn, robustness to (i) varying the signal-to-noise ratio, (ii) unequal error variances and (iii) time variation in the error covariance matrix (i.e. stochastic volatility).

{\it Signal-to-noise ratio.} In the paper, we consider ${\bf\Sigma}_u=0.01\cdot\mathbf{I}_{k}$, corresponding to a signal-to-noise ratio\footnote{Defined as the maximum eigenvalue of the parameter matrix over the maximum eigenvalue of the error covariance matrix.} of around 100. To investigate the sensitivity of the results to a lower signal-to-noise ratio, we re-ran the simulation study with ${\bf\Sigma}_u=0.1\cdot\mathbf{I}_{k}$, corresponding to a signal-to-noise ratio around 10 and ${\bf\Sigma}_u=\mathbf{I}_{k}$, corresponding to a signal-to-noise ratio around one.

{\it Unequal error variances.} We investigate whether the HLag methods behave comparably if one group of time series has a large residual variance and another group has a small residual variance. To this end, we consider one group (series 1 to 5) with  residual variance one, and the other group (series 6 to 10) with  residual variance equal to 0.5.

{\it Stochastic volatility.} As stochastic volatility is an important feature for macroeconomic forecasting \citep{Clark15}, we investigate the performance of all methods in the presence of parametric variation in the error covariance matrix. 
Note that none of the  methods considered in this paper account for stochastic volatility and, hence, their forecast accuracy is expected to suffer. Nevertheless, it remains interesting to investigate their sensitivity to the presence of parametric variation in the VAR errors.  

We consider the VAR-SV model of \cite{Clark15} which includes the conventional macroeconomic formulation of a random walk process for log volatility. In particular, we take
$$ 
\mathbf{u}_t = {\bf A}^{-1}\boldsymbol{\Lambda}_t^{0.5} \boldsymbol{\varepsilon}_t, 
$$
with $\boldsymbol{\varepsilon}_t \sim N({\bf 0}, {\bf I}_k)$ , ${\bf A} = \begin{bmatrix}
1 & {\bf 0} \\
{\bf 0.5} & {\bf I}_{k-1}
\end{bmatrix}$ 
and 
$ \  \boldsymbol{\Lambda}_t = \text{diag}(\lambda_{1,t}, \ldots, \lambda_{k, t})$ where
$$
\text{log}(\lambda_{i,t}) = \text{log}(\lambda_{i,t-1}) + v_{i,t},
$$
with $
v_{i,t} = (v_{1,t}, \ldots, v_{k,t})^\top \sim N({\bf 0}, 0.01\cdot {\bf I}_k)
$.

\begin{table}
	\resizebox{0.72\textwidth}{!}{\begin{minipage}{\textwidth}
			\begin{tabular}{llccccccc} \hline 
				Class 		& Method && SNR$\approx$100  && SNR$\approx$10 & SNR$\approx$1 & Unequal  & Stochastic \\ 
				& && (in paper) &&&&Variances& Volatility \\\hline 
				HLag 		& Componentwise 	&& 0.0125 (0.0003) && 0.1245 (0.0026) & 1.1814 (0.0228) & 0.9222 (0.0412) & 3.5548 (0.2148)\\
				& Own-other 					&& 0.0128 (0.0003) && 0.1278 (0.0027) & 1.2075 (0.0234) & 0.9414 (0.0421) & 3.6507 (0.2188)\\
				& Elementwise 					&& 0.0126 (0.0003) && 0.1262 (0.0026) & 1.2000 (0.0230) & 0.9414 (0.0421) & 3.6122 (0.2168)\\
				VAR     	& Lasso			 	&& 0.0131 (0.0003) && 0.1305 (0.0027) & 1.2365 (0.0236) & 0.9708 (0.0434) & 3.6818 (0.2196)\\
				& Lag-weighted lasso 			&& 0.0144 (0.0007) && 0.1439 (0.0066) & 1.5636 (0.1402) & 1.0803 (0.0483) & 4.2317 (0.2927)\\
				& Least squares AIC				&& 0.0136 (0.0003) && 0.1358 (0.0029) & 1.3250 (0.0261) & 1.0134 (0.0453) & 3.9678 (0.2260) \\
				& Least squares BIC				&& 0.0155 (0.0003) && 0.1554 (0.0034) & 1.5189 (0.0304) & 1.2491 (0.0559) & 3.9970 (0.2333) \\
				& VAR(1)					&& 0.0164 (0.0004) && 0.1643 (0.0035) & 1.5859 (0.0312) & 1.2555 (0.0276) & 4.4612 (0.2248) \\
				BVAR     	& BGR 				&& 0.0129 (0.0003) && 0.1295 (0.0027) & 1.2325 (0.0240) & 0.9688 (0.0433) & 3.6102 (0.2044)\\
				& GLP 							&& 0.0125 (0.0003) && 0.1253 (0.0026) & 1.2054 (0.0232) & 0.9312 (0.0194) & 3.6572 (0.2282)\\
				& CCM							&& 0.0129 (0.0003) && 0.1274 (0.0026) & 1.2128 (0.0236) & 0.9430 (0.0204) & 3.5280 (0.2088) \\
				Factor 		& DFM 				&& 0.0214 (0.0005) && 0.2142 (0.0054) & 1.9738 (0.0421) & 1.5231 (0.0395) & 4.6188 (0.2338)\\
				& FAVAR 						&& 0.0191 (0.0004) && 0.1913 (0.0043) & 1.7898 (0.0380) & 1.2990 (0.0293) & 4.3722 (0.2346)\\ 
				
				Other & AR						&& 0.0475 (0.0013) && 0.4753 (0.0134) & 4.5135 (0.1333) & 3.4895 (0.0987) & 11.9719 (0.6454)\\
				& Sample mean					&& 0.2067 (0.0083) && 2.0675 (0.0826)& 20.0255 (0.8383) & 14.5514 (0.6508) & 69.9780 (7.4943)\\
				& Random walk					&& 0.6268 (0.0256) && 6.2679 (0.2561)& 61.9335 (2.7009) & 44.8748 (2.0068) & 223.4107 (26.2932) \\ \hline 
			\end{tabular}
	\end{minipage} }
	\caption{Robustness to various choices of error covariance matrix: Out-of-sample  mean squared forecast error (standard errors are in parentheses).\label{covmatrix}}
\end{table}

Table \ref{covmatrix} gives the forecast performance of the methods under the various choices of error covariance matrix.
When decreasing the signal-to-noise ratio, the forecast accuracy of all methods decreases accordingly, as expected. 
Similarly, under unequal error variances and in the presence of stochastic volatility, the forecast accuracy  of all methods suffers compared to their performance in the original design (column 1). Importantly, the relative performance of the HLag methods to the other methods is, mainly, unaffected. One exception concerns the presence of stochastic volatility where even the homoscedastic BVAR of \cite{Carriero17}, which does not account for stochastic volatility, outperforms the HLag methods.  Their heteroskedastic BVAR, which accounts for stochastic volatility, is expected to perform even better in such settings.

\subsection{Relaxed VAR Estimation}\label{RVARappendix}
Since the lasso and its structured counterparts are known to shrink non-zero regression coefficients, in practice, they are often used for model selection, followed by refitting the reduced model using least squares \citep{relaxed}.  In this section, we detail our approach to refit based on the support selected by our procedures while taking into consideration both numerical stability as well as computational efficiency.

Let $\widehat{\PhiB}$ denote the coefficient matrix recovered from one of our sparsity-imposing algorithms (e.g. HLag, Lasso-VAR) and suppose that it contains $r$ nonzero coefficients.  In order to take the support recovered into account we introduce $\mathbf{V}$, a $k^{2}p\times r$ \emph{restriction matrix} of rank $r$ that denotes the location of nonzero elements in $\hat{\PhiB}$.  Defining $\beta$ as the vec of the nonzero entries of $\widehat{\PhiB}$, we obtain the relationship 
\begin{align*}
\text{vec}(\hat{\PhiB})=\mathbf{V}\beta.  
\end{align*}
We can then express the \emph{Relaxed Least Squares} estimator as:
\begin{align}
\label{RVAR}
\text{vec}(\widehat{\PhiB}_{\text{Relaxed}})=\mathbf{V}[\mathbf{V}^{\top}(\Z\Z^{\top}\otimes \utwi{I}_k)\mathbf{V}]^{-1}\mathbf{V}^{\top}(\Z\otimes \utwi{I}_k)\text{vec}(\Y),  
\end{align}
in which $\otimes$ denotes the Kronecker operator.  In general, it is ill-advised to directly form equation \eqref{RVAR}.  First, performing matrix operations with $\Z\otimes \utwi{I}_k$, which has dimension $kT\times k^2p$, can be very computationally demanding, especially if $k$ is large.  Second, in the event that $r\approx T$, the resulting estimator can be very poorly conditioned.  
To obviate these two concerns, we propose a slight adaptation of the techniques detailed in \cite{neumaier} that computes a variant of equation \eqref{RVAR} using a QR decomposition to avoid explicit matrix inversion.  Additionally, if the resulting matrix is found to be ill-conditioned, a small ridge penalty should be utilized to ensure numerically-stable solutions.

\subsection{Refinements}
As opposed to performing a Kronecker expansion we instead consider imposing the restrictions by row in $\widehat{\PhiB}$ and define $V_1,\dots,V_k$ as $kp\times r_i$ restriction matrices of rank $r_1,\dots,r_k$, denoting the number of nonzero elements in each row of $\PhiB$.  We can then calculate each row of $\widehat{\PhiB}_{\text{Relaxed}}$ by
\begin{align}
\label{RVAR2} \nonumber
\widehat{\PhiB}_{\text{Relaxed}_i}=\big(V_i(V_i^{\top}\Z\Z^{\top}V_i)^{-1}V_i^{\top}\Z\Y_{i}\big)^{\top}.  
\end{align}
Now, following \cite{neumaier}, construct the matrix $\mathbf{K}_i=[(V_i\Z)^{\top},\Y_i]$.  We then compute a QR factorization of $\mathbf{K}_i$
\begin{align*}
\mathbf{K}_i=QR,  
\end{align*}
in which Q is an orthogonal matrix and R is upper triangular of the form:
\begin{align*}
&R= \begin{blockarray}{c@{}cc@{\hspace{4pt}}cl}
& r_i & 1  & & \\
\begin{block}{[c@{\hspace{5pt}}cc@{\hspace{5pt}}c]l}
& R_{11} & R_{12}  & & \mLabel{r_i} \\
& 0 & R_{22}  & & \mLabel{T-r_i} \\
\end{block}
\end{blockarray}  
\end{align*}
As expanded upon in \cite{neumaier}, we can compute
\begin{align*}
&\widehat{\PhiB}_{\text{Relaxed}_i}=\big(V_iR_{12}^{\top}R_{11}(R_{11}^{\top}R_{11})^{-1}\big)^{\top},\\
=&\big(V_iR_{12}^{\top}R_{11}R_{11}^{-1}(R_{11}^{\top})^{-1}\big)^{\top},\\
=&\big(V_iR_{12}^{\top}(R_{11}^{\top})^{-1}\big)^{\top},\\
=&\big(V_i(R_{11}^{-1}R_{12})^{\top}\big)^{\top},  
\end{align*}
which can be evaluated with a triangular solver, hence does not require explicit matrix inversion.
In the event that $\mathbf{K}$ is poorly conditioned, to improve numerical stability, we add a small ridge penalty.  It is suggested by \cite{neumaier} to add a penalty corresponding to scaling a diagonal matrix D consisting of the Euclidean norms of the columns of $\mathbf{K}$ by $(r_i^2+r_i+1)\epsilon_{\text{machine}}$, in which $\epsilon_{\text{machine}}$ denotes machine precision.  The full refitting algorithm is detailed in Algorithm \ref{ALGRLS}.

\begin{algorithm}[h]
	\caption{\label{ALGRLS} Relaxed Least Squares }
	\begin{algorithmic}
		\Require $\Z,\Y,V_1,\dots,V_k$  
		\For{$i=1,2,\dots,k$}
		\State $\mathbf{K}_i\leftarrow [(V_i\Z)^{\top},Y_i]$
		\State $D\leftarrow (r_i^2+r_i+1)\epsilon_{\text{machine}}\text{diag}(\|\mathbf{K}_{i\cdot}\|_2)$
		\State $R,Q\leftarrow QR(
		\begin{bmatrix}
		\mathbf{K}_i  \\
		D
		\end{bmatrix}
		)$
		\State $\widehat{\PhiB}_{\text{Relaxed}_i}\leftarrow \big(V_i(R_{11}^{-1}R_{12})^{\top}\big)^{\top}$
		\EndFor\\
		\Return $\widehat{\PhiB}_{\text{Relaxed}}$.
	\end{algorithmic}
\end{algorithm}

\section{Stock and Watson Application} \label{SWAppendix}
\begin{table}
	\resizebox{0.83\textwidth}{!}{\begin{minipage}{\textwidth}
			\begin{tabular}{lllc} \hline 
				Group & Brief description & Examples of series & Number  \\ 
				&  &   & of series \\\hline 
				1 & GDP components & GDP, consumption, investment & 20 \\
				2 & IP & IP, capacity utilization & 15 \\
				3 & Employment & Sectoral and total employment and hours & 20 \\
				4 & Unemployment rate & Unemployment rate, total and by duration & 7 \\
				5 & Housing & Housing starts, total and by region & 6 \\
				6 & Inventories & NAPM inventories, new orders & 6 \\
				7 & Prices & Price indexes, aggregate and disaggregate; commodity prices & 52 \\
				8 & Wages & Average hourly earnings, unit labor cost & 9 \\
				9 & Interest rates & Treasuries, corporate, term spreads, public-private spreads & 14 \\
				10 & Money & M1, M2, business loans, consumer credit & 8 \\
				11 & Exchange rates & Average and selected trading partners & 5 \\
				12 & Stock prices & Various stock price indexes & 5 \\
				13 & Consumer expectations & Michigan consumer expectations & 1 \\ \hline 
			\end{tabular}
	\end{minipage} }
	\caption{Macroeconomic categories of series in the 168-variable data set, following the classification of \cite{stock12} their Table 1. \label{SWgroups}}	
\end{table}

To make the $k=168$ variables of the Stock and Watson data approximately stationary, we apply the transformation codes provided by \cite{stockdataset}.
A brief description of each variable, along with the transformation code to make them approximately stationary can be found in the Data Appendix of \cite{koop}.

\begin{figure}
	\includegraphics[width=\textwidth]{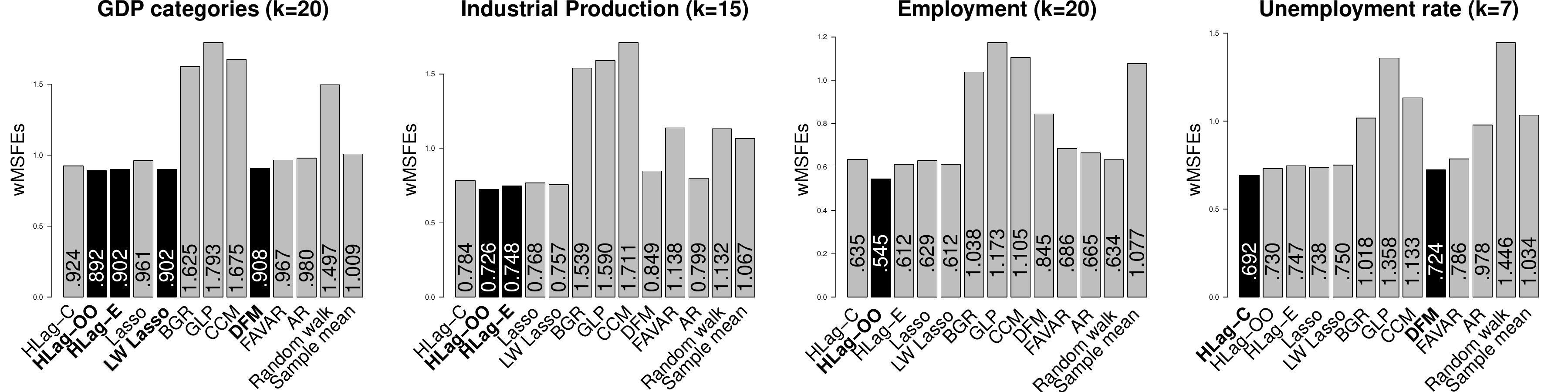}                
	\includegraphics[width=\textwidth]{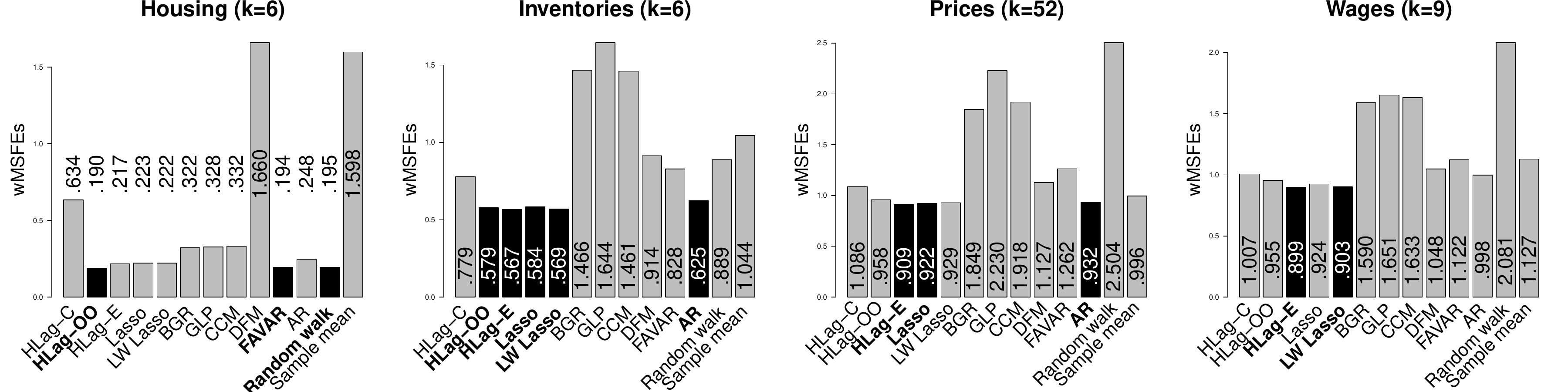}                
	\includegraphics[width=\textwidth]{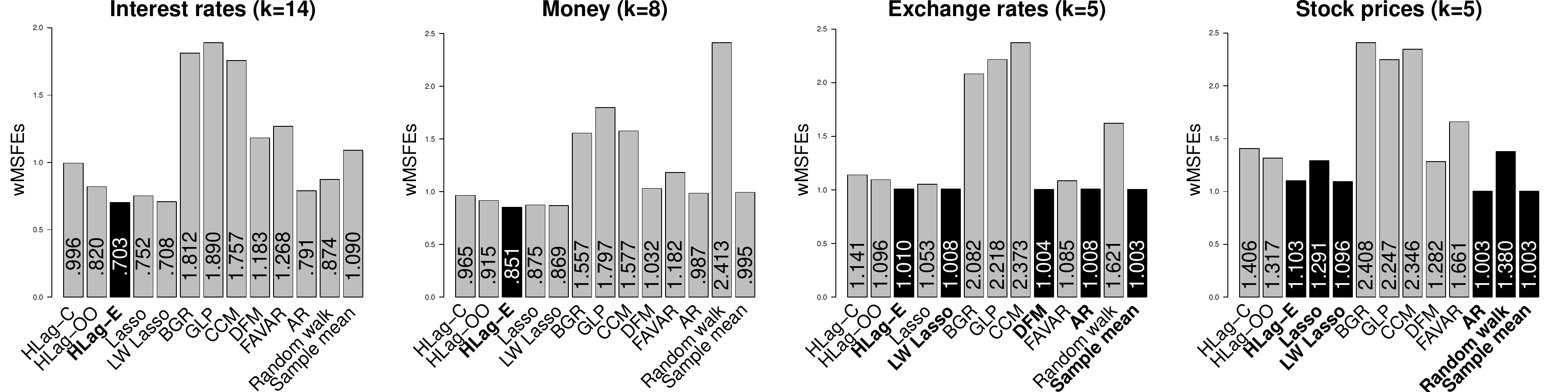}                
	\includegraphics[width=\textwidth]{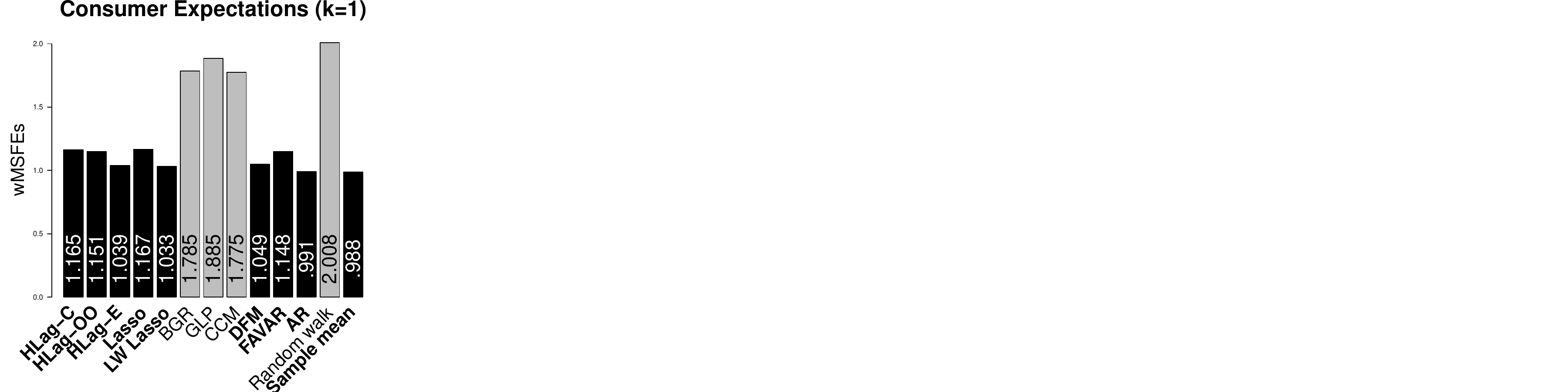} 
	\caption{Rolling out-of-sample one-step ahead $wMSFE$ of different categories of macroeconomic indicators in the \textit{Large} VAR. For each category, forecast methods in the 75\% MCS are in black. \label{wMSFEcategories}}
\end{figure}

All $168$ variables are classified into one of 13 macroeconomic categories, detailed in Table \ref{SWgroups}. The good performance of the HLag methods across all variables is confirmed by a sub-analysis on the 13 macroeconomic categories. 
Figure \ref{wMSFEcategories} breaks down the results of the \textit{Large} VAR by the 13 macroeconomic categories. Generally speaking, the flexible elementwise HLag is the best performing forecasting method; for 10 out of 13 categories, it is included in the MCS. The second best performing methods are own-other HLag and the lag-weighted lasso (both for 6 out of 13 categories in the MCS).

Upon examination of the different categories, three groups can be distinguished.
The first group consists of categories with a single preferred forecast method, always an HLag method. Elementwise HLag is preferred  for interest rates and money;  own-other HLag for employment series. 
The second group consists of categories with several, but a limited number (between 2 and 4) of preferred  methods. Series in the second group are major measures of real economic activity (GDP components, industrial production, unemployment rate, prices), housing, and wages. The strong performance of elementwise and own-other HLag is re-confirmed in the majority of cases (3 out of 5 categories), but the MCS is extended by the lag-weighted lasso and DFM (2 out of 5 categories), or componentwise HLag, FAVAR and random walk (1 out of 5 categories). 
The third group consists of categories which have a larger number of preferred forecast methods, like inventories and hard-to-predict series such as exchange rates, stock prices and consumer expectations. For the latter categories, in line with \cite{stock12}, we find multivariate forecast methods to provide no meaningful reductions over simple univariate methods (AR  or sample mean).

\begin{figure}
	\includegraphics[width=\textwidth]{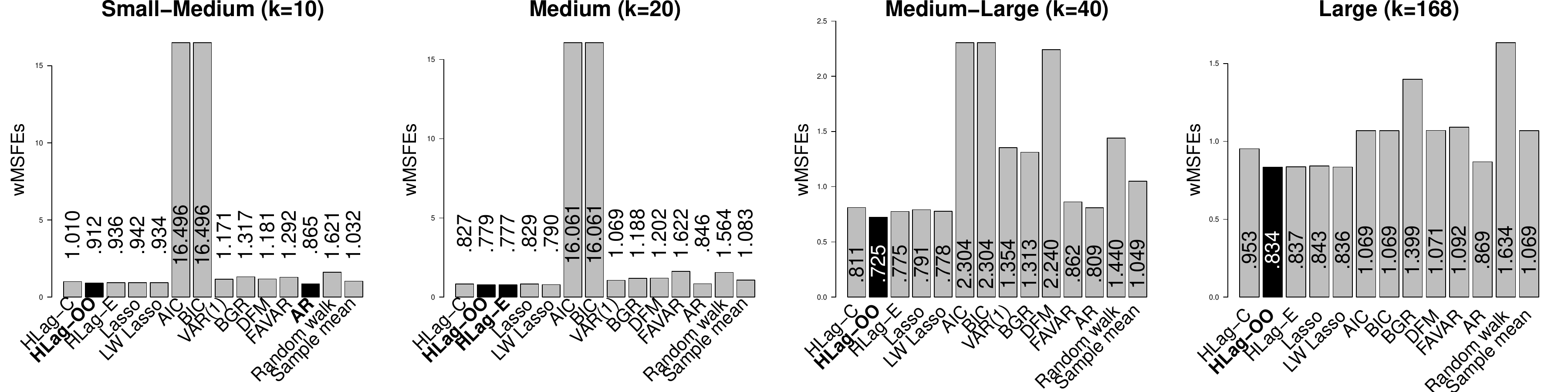} 
	\caption{\label{h1wMSFEallpmax13} Rolling out-of-sample one-step-ahead $wMSFE$ for the four VAR sizes with $pmax = 13$. For each VAR size, forecast methods in the 75\% Model Confidence Set are in black.}
\end{figure}

\begin{figure}
	\includegraphics[width=\textwidth]{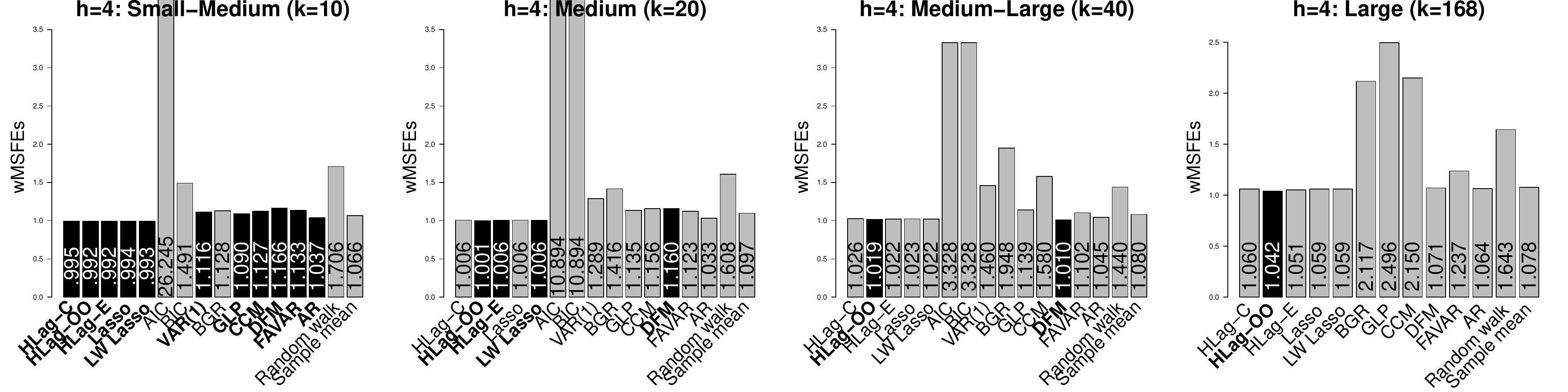}                
	\includegraphics[width=\textwidth]{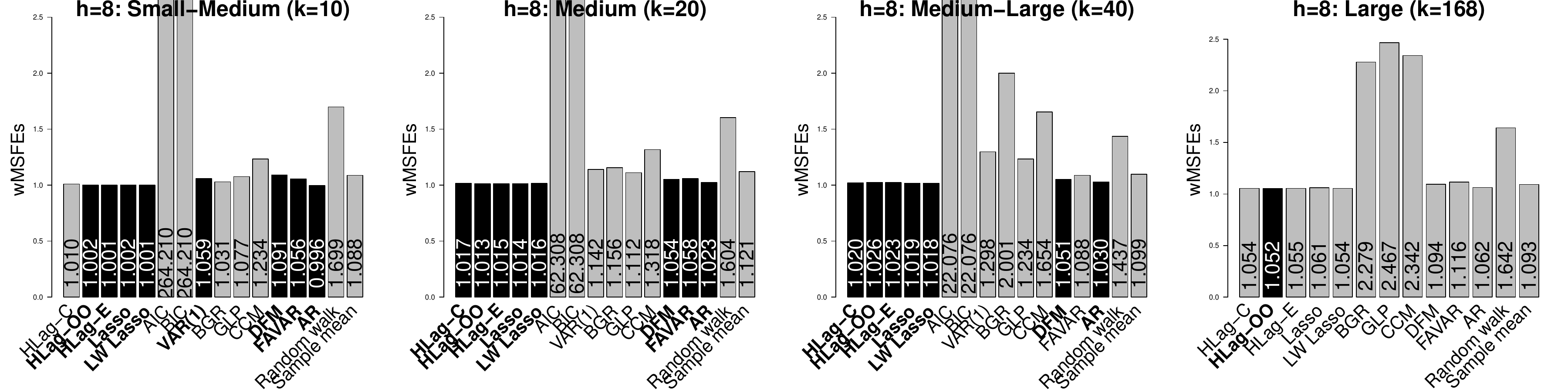} 
	\caption{\label{h4-8-wMSFEall} Rolling out-of-sample one-step-ahead $wMSFE$
		for the four VAR sizes at forecast horizon $h=4$ (top) and $h=8$ (bottom). For each VAR size, forecast methods in the 75\% MCS are in black.}
\end{figure}

\begin{figure}
	\includegraphics[width = \textwidth]{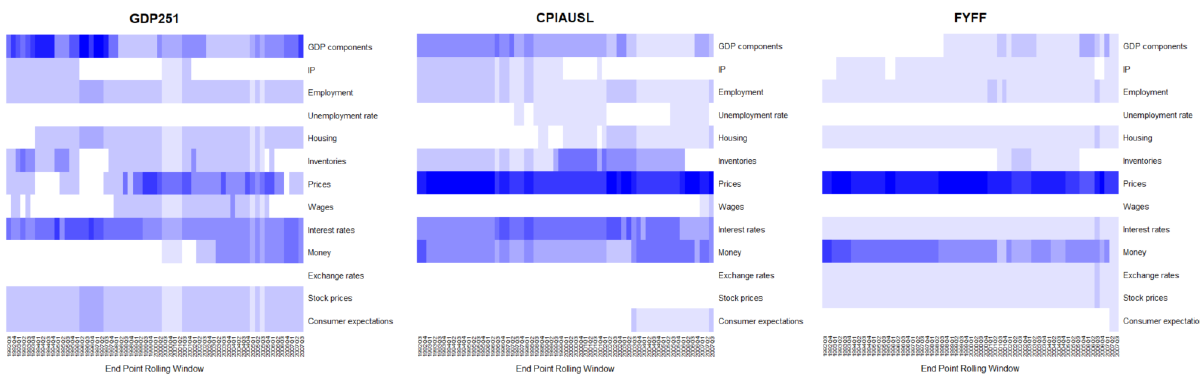} 
	\caption{Fraction of non-zero coefficients in each of the 13 macro-economic categories to the total number of non-zero coefficients in the \textit{Medium-Large} VAR  estimated by elementwise HLag when forecasting \textit{GDP251} (GDP growth, left), \textit{CPIAUSL} (inflation, middle) and \textit{FYFF} (Federal Funds Rate, right). The horizontal axis represents the ending date of a rolling window.}\label{plotsHLagE}
\end{figure}

Results on additional sensitivity analyses concerning the choice of the maximal lag order pmax and  forecasts horizon are provided in Figures \ref{h1wMSFEallpmax13} and \ref{h4-8-wMSFEall} respectively. Results on the stability of the lag selection results are displayed in Figure \ref{plotsHLagE}.

We focus on the Stock and Watson macroeconomic data set since it is readily available and popular in the literature on macroeconomic forecasting. A more recent variant is available as the FRED-QD data set, a quarterly version of the Federal Reserve Economic Data database introduced in \cite{McCracken2016}. 
We have performed the same empirical analysis on the FRED-QD containing $k=210$ variables from Quarter 3, 1959 to Quarter 4, 2018 ($T=238$). Similar findings are obtained: 
(i) Own-other and elementwise HLag perform comparable to the lasso methods and AR for small VAR sizes, but outperform all others for the \textit{Large} VAR and a short forecast horizon.  
(ii) Own-other HLag is the preferred forecast method for several major macroeconomic indicators such as national income and product accounts and industrial production. For difficult to predict  indicators, such as exchange rates, gains over the AR model are difficult to attain.

%\clearpage
\section{Financial Application} \label{finance_appendix}
The financial data set contains information on the realized variances of $k=16$ stock market indices listed in Table \ref{finance_descr}. All time series are log-transformed to make them stationary. 

\begin{table}[h]
	\centering
	\begin{tabular}{lll} \hline
		Variable & Description &  \\ \hline
		AEX&Amsterdam Exchange Index & \\
		AORD& All Ordinaries Index& \\ 
		BFX& Belgium Bell 20 Index & \\   
		BVSP& BOVESPA Index & \\ 
		DJI& Dow Jones Industrial Average & \\ 
		FCHI& Cotation Assist\'ee en Continu Index & \\ 
		FTSE& Financial Times Stock Exchange Index 100 & \\ 
		GDAXI& Deutscher Aktienindex & \\ 
		HSI& HANG SENG Index & \\ 
		IXIC& Nasdaq stock index & \\ 
		KS11& Korea Composite Stock Price Index & \\ 
		MXX& IPC Mexico & \\ 
		RUT& Russel 2000 & \\ 
		SPX& Standard \& Poor's 500 market index & \\ 
		SSMI& Swiss market index & \\ 
		STOXX50E&  EURO STOXX 50 & \\ \hline 
	\end{tabular}
	\caption{Variables used in the financial application. \label{finance_descr}}
\end{table}

\section{Energy Application} \label{energy_appendix}

The energy data set contains information on $k=26$ variables. A brief description of each variable, taken from https://archive.ics.uci.edu/ml/data sets/Appliances+energy+prediction, is provided in Table \ref{energy_descr}, along with the transformation code to make it approximately stationary. The transformation codes are: 1 = first difference of logged variables, 2 = first difference.

\begin{table}[h]	\centering
	\begin{tabular}{lll} \hline
		Variable & Description & Code \\ \hline
		Appliances & energy use in Wh & 1 \\ 
		Lights &  energy use of light fixtures in the house in Wh&2\\
		T1&  Temperature in kitchen area, in Celsius& 1 \\ 
		RH1& Humidity in kitchen area, in \% & 1 \\ 
		T2&  Temperature in living room area, in Celsius& 1 \\ 
		RH2&  Humidity in living room area, in \%& 1 \\ 
		T3&  Temperature in laundry room area& 1 \\ 
		RH3 &  Humidity in laundry room area, in \%& 1 \\ 
		T4 & Temperature in office room, in Celsius & 1 \\ 
		RH4& Humidity in office room, in \% & 1 \\ 
		T5& Temperature in bathroom, in Celsius & 1 \\ 
		RH5& Humidity in bathroom, in \% & 1 \\ 
		T6& Temperature outside the building (north side), in Celsius &2\\
		RH6& Humidity outside the building (north side), in \% & 1 \\ 
		T7 & Temperature in ironing room , in Celsius & 1 \\ 
		RH7& Humidity in ironing room, in \% & 1 \\ 
		T8&  Temperature in teenager room 2, in Celsius& 1 \\ 
		RH8&  Humidity in teenager room 2, in \%& 1 \\ 
		T9& Temperature in parents room, in Celsius & 1 \\ 
		RH9& Humidity in parents room, in\% & 1 \\ 
		To& Temperature outside (from Chievres weather station), in Celsius &2\\
		Pressure& From Chievres weather station, in mm Hg & 1 \\ 
		RHout& Humidity outside (from Chievres weather station), in \% & 1 \\ 
		Wind speed & From Chievres weather station  in m/s &2\\
		Visibility&   From Chievres weather station, in km & 1 \\ 
		Tdewpoint&  From Chievres weather station, Â°C &2\\ \hline
	\end{tabular}
	\caption{Variables used in the energy application. \label{energy_descr}}
\end{table}

\spacingset{1} 
\bibliography{harbib}
\spacingset{1.5} % Double spacing
	
\end{document}